\documentclass[journal=mamobx,manuscript=article]{achemso}

\usepackage{chemformula} % Formula subscripts using \ch{}
\usepackage[T1]{fontenc} % Use modern font encodings
\usepackage[utf8]{inputenc}
\DeclareUnicodeCharacter{0308}{\"}
\usepackage{amssymb}
\usepackage{dcolumn}% Align table columns on decimal point
\usepackage{multirow}
\usepackage{float}
\usepackage{booktabs}
\usepackage{hyperref}
\usepackage{cleveref}
\usepackage{mathrsfs,amsmath,bm}
\usepackage{color}
\usepackage{xr}
\newcommand{\bn}{\bm{n}}

\newcommand{\bk}{\bm{k}}
\newcommand{\br}{\bm{r}}
\newcommand{\bR}{\bm{R}}
\newcommand{\bP}{\bm{P}}
\newcommand{\bJ}{\bm{J}}

\author{Jing Chen}
\affiliation[Xiangtan University]
{Hunan Key Laboratory for Computation and Simulation in Science and Engineering, Key Laboratory of Intelligent Computing and Information Processing of Ministry of Education, School of Mathematics and Computational Science, Xiangtan University, Xiangtan, Hunan, China, 411105}
\altaffiliation{Jing Chen and Aiping Zhu contributed equally to this work}

\author{Aiping Zhu}
\affiliation[Xiangtan University]
{Hunan Key Laboratory for Computation and Simulation in Science and Engineering, Key Laboratory of Intelligent Computing and Information Processing of Ministry of Education, School of Mathematics and Computational Science, Xiangtan University, Xiangtan, Hunan, China, 411105}
\altaffiliation{Jing Chen and Aiping Zhu contributed equally to this work}

\author{Dan Wei}
\affiliation[Xiangtan University]
{Hunan Key Laboratory for Computation and Simulation in Science and Engineering, Key Laboratory of Intelligent Computing and Information Processing of Ministry of Education, School of Mathematics and Computational Science, Xiangtan University, Xiangtan, Hunan, China, 411105}
\email{danwei@xtu.edu.cn}

\author{An-Chang Shi}
\affiliation[McMaster University]
{Department of Physics and Astronomy, McMaster University, Hamilton, Ontario, L8S4M1, Canada}
\email{shi@mcmaster.ca}

\author{Kai Jiang}
\affiliation[Wuhan University]
{Wuhan Institute for Math \& AI, Wuhan University, Wuhan, Hubei, China, 430072}
\email{kaijiang@whu.edu.cn}

\title
  {Structure and energetics of grain boundaries in self-assembled double-gyroid block copolymer networks}
 
\begin{document}
	\begin{figure}[!htbp]
		\centering
		\includegraphics[width=1\linewidth]{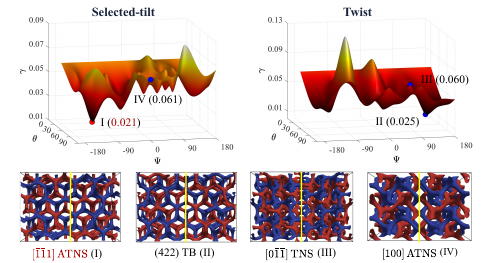}
		\caption*{Graphic Table of Contents.}
		%\label{for Table of Contents use only}
	\end{figure}
\newpage
\begin{abstract}
Grain boundaries (GBs) are ubiquitous defects in crystalline materials, but remain less explored in block copolymer ordered phases. Here, we develop a self-consistent field theory framework to investigate GB structure and energetics in double-gyroid (DG) diblock copolymer networks. Two energy landscapes are obtained for tilt and twist GBs as functions of GB orientations revealing multiple local minima. The lowest-energy GB among the sampled GBs is an asymmetric-tilt network-switching GB ($[\bar{1}\bar{1}1]$ ATNS) rather than the experimentally observed $(422)$ twin boundary (TB) among the sampled GBs. Comparative analyses of tilt ($[\bar{1}\bar{1}1]$, $[100]$ ATNS) and twist ($(422)$ TB, $[0\bar{1}\bar{1}]$ TNS) GBs reveal that, unlike enthalpy-dominated hard matter, GB stability in DG networks is entropy-driven. Twist-type GBs generate new nodes and disrupt nodal coplanarity, causing chain-packing frustration and entropy penalties. Although the [100] ATNS does not generate new nodes, its disruption of nodal coplanarity still incurs large entropy penalties. Conversely, the $[\bar{1}\bar{1}1]$ ATNS preserves favorable network connectivity and minimizes conformational constraints, making it the energetically preferred GB.
\end{abstract}

\section{Introduction}
Block copolymers (BCPs) can self-assemble into various ordered structures via microphase separation of chemically immiscible blocks\,\cite{bates1999block}. Beyond the classical lamellar, cylindrical, and spherical phases\,\cite{bates1999block}, BCPs are also known to form complex three-dimensional network structures, including the double gyroid (DG)\,\cite{hajduk1994gyroid, roy2011double}, double diamond (DD)\,\cite{chang2021mesoscale, asai2017tricontinuous}, and Fddd phases\,\cite{bailey2002noncubic, epps2004network}. These network phases have attracted sustained interest due to their potential applications in photonic crystals\,\cite{maldovan2002photonic, urbas2002bicontinuous, saba2011circular}, ion transport media\,\cite{goto2009development, prasad2018anatomy} and plasmonic metamaterials\,\cite{oh2012origin, lee201425th, hur2011three}. However, in practical applications, BCPs rarely form ideal single-crystalline structures.
Inevitable impurities during synthesis, together with kinetic limitations during microphase separation and grain growth, lead to the formation of various defects\,\cite{li2015defects, li2014linking, feng2019topological, feng2019seeing, miyata2022dislocation, shan2024nature}.
Among these, grain boundaries (GBs) are ubiquitous, which are interfaces separating grains of the same phase with different orientations. 
Their presence disrupts the structural symmetry, and can strongly influence the macroscopic properties of materials.

In hard matter such as metals and alloys, the energetics of GBs is primarily governed by inter-atomic interactions\,\cite{gottstein2009grain, murdoch2013estimation}. Strong interatomic bonding and large elastic moduli impose stringent lattice constraints, so that even angstrom-scale atomic displacements from ideal lattice sites incur substantial energetic penalties\,\cite{read1950dislocation}. As a consequence, GB energies are primarily determined by geometric lattice mismatch and elastic distortion\,\cite{sutton1995interfaces}. High-symmetry twin boundaries (TBs), which minimize atomic misfit and strain, emerge as the representative low-energy configurations in hard matter\,\cite{bulatov2014grain, tschopp2015symmetric, hahn2016symmetric, li2019atomistic}.
In contrast, the networked phases of BCPs form on nanometer-to-micrometer length scales.
Their stability originates not from strong chemical bonding, but rather from a subtle competition between weak non-covalent interactions (e.g., van der Waals forces) and the conformational entropy of the polymer chains. 
The intrinsic plasticity of ``soft'' lattices affords BCP networks high geometric deformability, continuous skeletal connectivity, and significant topological freedom. During GB formation, geometric incompatibility and elastic penalties are mitigated by long relaxation times, chain conformational adjustments, and network reconfigurations, rather than being constrained by strict lattice matching. Consequently, the energy landscape of GBs in BCP networked phases is fundamentally reshaped. Symmetry-based selection rules, traditionally applied to hard matter, may no longer suffice as a standalone metric for identifying low-energy GB configurations. 
The flexibility of GBs in soft matter raises a fundamental question: beyond the experimentally observed TBs, do other configurations with even lower GB energies exist, and what physical mechanisms govern their stability in BCP network phases? To date, these questions remain largely unexplored.

Experimentally, early studies of GBs in BCPs primarily focused on lamellar phases, where twist and tilt GBs were characterized via transmission electron microscopy\,\cite{gido1993lamellar, gido1994lamellar1, carvalho1995grain, jinnai2006direct, ryu2013role}. 
In contrast, studies of GBs in complex networked phases remain limited, largely centered on the structural characterization and geometric analysis of TBs in DG and DD phases\,\cite{han2011spontaneous, han2020crystal, liu2021symmetry,feng2023soft,vignolini2011a, feng2021visualizing}. Non-twin GBs in these networked structures have received far less attention and remain largely unexplored.
Theoretically, previous studies of network-phase GBs have primarily relied on minimal surface models\,\cite{chen2018minimal, han2020crystal} or Landau-type theories\,\cite{belushkin2009twist, chen2026computational}. While these approaches effectively capture geometric features, their application to BCP systems is not straightforward. 
In particular, our recent study developed a computational framework for evaluating the energies of DG GBs within a phase-field model and applied it to four GB structures, including the experimentally observed $(422)$ TB, an asymmetric-tilt network-switching GB (ATNS), and two high-energy tilt GBs\,\cite{chen2026computational}. The calculations showed that the ATNS possesses a lower GB energy than the experimentally observed $(422)$ TB, and geometric analyses suggested that nodal coplanarity may play an important role in determining the GB energy. However, owing to the phenomenological nature of the Landau--Brazovskii (LB) model, 
it is difficult to provide a molecular-level interpretation of the energetic ordering in terms of polymer chain conformations and configurational entropy, or to establish a direct correspondence with specific block copolymer systems.
Consequently, the microscopic thermodynamic mechanism governing GB stability of network phases remains unclear.

For inhomogeneous polymeric systems, the self-consistent field theory (SCFT) provides a powerful framework and offers the distinct advantage of simultaneously capturing polymer chain conformations, entropy and interaction energies with well-defined physical parameters\,\cite{fredrickson2006equilibrium}. Although SCFT has been extensively applied to periodic structures\,\cite{bates1999block, matsen1994stable, zheng1995morphology, matsen2002standard,jiang2024adaptive} and quasicrystals\,\cite{duan2018stability, wang2025stability}, its application to the study of GBs remains notably rare. To date, SCFT-based GB research is restricted to simple lamellar phases\,\cite{gido1994lamellar, gido1994lamellar1, gido1997lamellar, hashimoto1994ordered, matsen1997kink, duque2000self}, whereas the GBs in complex networked phases have yet to be explored.

In this study, we perform a systematic SCFT investigation of GBs in the DG phase of BCPs, where the DG phase consists of two interpenetrating but disconnected networks of opposite chirality.
Our aim is to understand the structures and stability mechanisms of DG GBs.
By developing a SCFT framework tailored for GB computations, we construct two GB energy landscapes in the orientation space $(\theta,\Psi)$ for tilt and twist GBs, respectively, where each point represents a distinct relative orientation between the two grains.
These landscapes capture rich GB structures and multiple local minima within both tilt and twist configurations, enabling the identification of low-energy GBs. Interestingly, our results reveal that the experimentally observed $(422)$ TB is not the lowest-energy GB among the sampled configurations. Instead, an asymmetric-tilt network-switching GB ($[\bar{1}\bar{1}1]$ ATNS) exhibits the lowest energy. The relative stability of the $[\bar{1}\bar{1}1]$ ATNS and $(422)$ TB is consistent with that found in the LB study\,\cite{chen2026computational}.
To elucidate the underlying mechanism, we investigate a representative set of GBs, including the low-energy $[\bar{1}\bar{1}1]$ ATNS and $(422)$ TB, together with higher-energy twist ($[0\bar{1}\bar{1}]$ TNS) and asymmetric-tilt ($[100]$ ATNS) GBs for comparison. Analyses of entropy and interaction energies, chain conformations, network topologies, and geometric deformations demonstrate that, GB stability in BCP network phases is primarily governed by entropy penalties associated with node formation and coplanarity, in stark contrast to the enthalpy-dominated behavior typically found in hard materials.

\section{Theoretical Framework}\label{sec:theory}
\begin{figure}[htbp]
	\centering
	\includegraphics[width=0.9\linewidth]{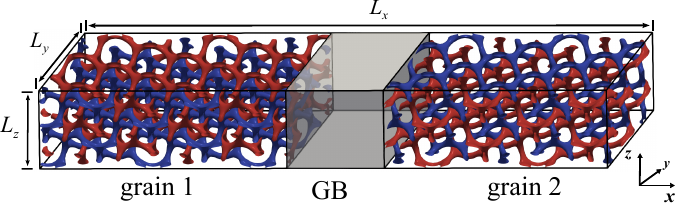}
	\caption{Schematic of the computational setup.}
	\label{framework0}
\end{figure}

This work focuses on the spontaneous formation of a GB when two grains of the same phase with different orientations come into contact. Our computational framework is illustrated schematically in Figure\,\ref{framework0}. The two grains, each with a predefined orientation, are positioned in the two opposing half-spaces, grain 1 at $x\leq -L_x/2$ and grain 2 at $x\geq L_x/2$. The intermediate region $-L_x/2< x < L_x/2$ is made sufficiently large to allow for the full relaxation of the GB. SCFT calculations are performed in the rectangular box of volume $V=L_x\times L_y \times L_z$. To fix the bulk structures corresponding to the two grain orientations at the box ends, anchoring boundary conditions are imposed at $x=\pm L_x/2$, while periodic boundary conditions are applied in the $y$- and $z$-directions. Under this setup, the grains on both sides evolve toward each other along the $x$-direction. Due to the difference of crystallographic orientations, they cannot achieve a defect-free connection upon meeting, resulting in the formation of a GB within a transition zone indicated by the gray region in Figure\,\ref{framework0}, where the network topology, chain conformation, and local geometric structure undergo a synergistic rearrangement to accommodate the orientation mismatch.

\subsection{SCFT model for GBs} 
We consider an incompressible melt of $n$ flexible AB diblock copolymers in the rectangular box.
Each polymer chain consists of one A block and one B block with the same statistical segment length.
The degree of polymerization of the A-block is denoted by $N_{A}=fN$, where $f$ represents the volume fraction of the A-block, and $N$ is the total degree of polymerization of the chain. The degree of polymerization of the B-block is given by $N_{B}=(1-f)N$. Following the standard SCFT framework for flexible AB diblock copolymers, which has been widely applied to investigate periodic ordered morphologies and phase behavior,\,\cite{helfand1975block,leibler1980theory,semenov1985contribution,ohta1986equilibrium,matsen1994stable,matsen2002standard,fredrickson2006equilibrium,li2006self} the flexible blocks are modeled as Gaussian chains, and their propagators obey the modified diffusion equations (MDEs),
\begin{equation}
    \begin{aligned}
         \frac{\partial}{\partial s}q(\br, s)&=\nabla^{2}_{\br}q(\br, s)-w_{\mathrm{A}}(\br)q(\br, s), \quad 0\le s\le f,\\
          \frac{\partial}{\partial s}q(\br, s)&=\nabla^{2}_{\br}q(\br, s)-w_{\mathrm{B}}(\br)q(\br, s),\quad f\le s\le 1,
    \end{aligned}
    \label{eqs3}
\end{equation}
where $q(\br,s)$ represents the probability of the $s$-th chain segment at spatial position $\br$ in mean fields $w_{\alpha}(\br),\,\alpha\in\{\mathrm{A}, \mathrm{B}\}$. 
The backward propagator $q^{\dagger}(\br, s)$ is defined analogously, satisfying similar MDEs,
\begin{equation}
    \begin{aligned}
          \frac{\partial}{\partial s}q^{\dagger}(\br, s) &= \nabla^{2}_{\br}q^{\dagger}(\br, s)-w_{\mathrm{B}}(\br)q^{\dagger}(\br, s), \quad  0\le s\le 1-f,\\
         \frac{\partial}{\partial s}q^{\dagger}(\br, s) &= \nabla^{2}_{\br}q^{\dagger}(\br, s)-w_{\mathrm{A}}(\br)q^{\dagger}(\br, s), \quad 1-f\le s\le 1.
    \end{aligned}
    \label{eqs30}
\end{equation}

To fix the crystallographic orientations of the two grains, inhomogeneous Dirichlet boundary conditions are imposed on the propagators along the $x$-direction. Meanwhile, periodic boundary conditions are applied in the $y$- and $z$-directions. 
Therefore, the boundary and initial conditions are given by,
\begin{equation}
    \begin{aligned}
        q(-L_x/2,y,z,s) &= q_{\mathrm{grain 1}}(-L_x/2,y,z,s),\\
        q(L_x/2,y,z,s) &= q_{\mathrm{grain 2}}(L_x/2,y,z,s),\\
        q(\br,0) &= 1,\\
        q^{\dagger}(-L_x/2,y,z,s) &= q^{\dagger}_{\mathrm{grain 1}}(-L_x/2,y,z,s),\\
        q^{\dagger}(L_x/2,y,z,s) &= q^{\dagger}_{\mathrm{grain 2}}(L_x/2,y,z,s),\\
        q^{\dagger}(\br, 0) &= 1.
    \end{aligned}
    \label{eqs300}
\end{equation}

Using these propagators, the single-chain partition function $Q$ and segment densities $\phi_{\alpha,\,\mathrm{GB}}(\br)$ ($\alpha\in\{\mathrm{A},\,\mathrm{B}\}$) of GB structures can be calculated as,
\begin{equation}
    \begin{aligned}
        &Q=\frac{1}{V}\int_{V} q(\br, s)q^\dagger(\br, 1-s)\,\mathrm{d}\br,\quad \forall s \in [0, 1],\\
        &\phi_{\mathrm{A},\, \mathrm{GB}}(\br)=\frac{1}{Q}\int_{0}^{f}q(\br,s)q^{\dagger}(\br,1-s)\,\mathrm{d}s,\\
        &\phi_{\mathrm{B},\, \mathrm{GB}}(\br)=\frac{1}{Q}\int_{f}^{1}q(\br,s)q^{\dagger}(\br,1-s)\,\mathrm{d}s.
    \end{aligned}
    \label{eqs2}
\end{equation}

The free energy per chain in units of $k_{\mathrm{B}} T$ is obtained by\,\cite{fredrickson2006equilibrium, jiang2015analytic}
\begin{equation}
    \begin{aligned}
    \frac{F_\mathrm{GB}}{n k_{\mathrm{B}} T} &= \frac{1}{V} \int_{V} \{ \chi N \phi_{\mathrm{A},\,\mathrm{GB}}(\br) \phi_{\mathrm{B},\,\mathrm{GB}}(\br) - \sum_{\alpha\in\{ \mathrm{A},\,\mathrm{B}\}}w_{\alpha}(\br) \phi_{\alpha,\,\mathrm{GB}}(\br) \\
    &+ \eta(\br) [1 -\sum_{\alpha\in\{ \mathrm{A},\, \mathrm{B}\}}\phi_{\alpha,\,\mathrm{GB}}(\br) ] \} \,\mathrm{d}\br -\ln Q,
    \end{aligned}
    \label{free_energy}
\end{equation}
where $T$ is the temperature, $k_\mathrm{B}$ is the Boltzmann constant, $\chi$ is the Flory-Huggins interaction parameter, and $\eta(\br)$ is the ``pressure'' field that ensures the incompressibility of the system. The free energy per chain $F_\mathrm{GB}/n k_{\mathrm{B}} T$ can be divided into two parts, the segment-segment interaction energy $F^{\mathrm{inter}}_\mathrm{GB}/nk_{\mathrm{B}}T$ and the entropic contribution energy $-TS_\mathrm{GB}/nk_{\mathrm{B}}T$,
\begin{equation}
    \begin{aligned}
    \frac{F^{\mathrm{inter}}_\mathrm{GB}}{nk_{\mathrm{B}}T} &= \frac{1}{V} \int_{V} \{ \chi N \phi_{\mathrm{A},\,\mathrm{GB}}(\br) \phi_{\mathrm{B},\,\mathrm{GB}}(\br) - \sum_{\alpha\in\{ \mathrm{A},\,\mathrm{B}\}}w_{\alpha}(\br) \phi_{\alpha,\,\mathrm{GB}}(\br) \\
    &+ \eta(\br) [1 -\sum_{\alpha\in\{ \mathrm{A},\, \mathrm{B}\}}\phi_{\alpha,\,\mathrm{GB}}(\br) ] \} \,\mathrm{d}\br, \\
    -\frac{TS_\mathrm{GB}}{nk_{\mathrm{B}}T} &= -\ln Q.
    \end{aligned}
\end{equation}

Setting the first-order variations of the free-energy functional with respect to $\eta(\br)$, $\phi_{\mathrm{A},\,\mathrm{GB}}(\br)$ and $\phi_{\mathrm{B},\,\mathrm{GB}}(\br)$ to zero gives the SCFT equations,
\begin{equation}
    \begin{aligned}
         \phi _{\mathrm{A},\,\mathrm{GB}}&( \br ) + \phi_{\mathrm{B},\,\mathrm{GB} }( \br ) = 1,\\
         w_{\mathrm{A}}(\br)&=\chi N\phi_{\mathrm{B},\,\mathrm{GB}}(\br)+\eta(\br),\\
         w_{\mathrm{B}}(\br)&=\chi N\phi_{\mathrm{A},\,\mathrm{GB}}(\br)+\eta(\br).\\
    \end{aligned}
    \label{eqs4}
\end{equation}

\subsection{Numerical Methods}

Starting from an initial GB configuration, the SCFT equations are solved iteratively to obtain the GB structures.
Each iteration involves solving the MDEs (eqs.\,\ref{eqs3}-\ref{eqs30}) to obtain the propagators, which are then used to compute the single-chain partition function, segment densities (eq.\,\ref{eqs2}), free energy (eq.\,\ref{free_energy}), and update the mean fields (eq.\,\ref{eqs4}). 
This process is repeated until the free energy difference between two consecutive iterations falls below a prescribed tolerance. 
 
The initial GB configuration is constructed by joining two grains with different orientations across the $x=0$ plane. Specifically, the segment density of the A-block in the bulk phase,  $\phi_{\mathrm{A},\,\mathrm{bulk}}(\br)$, is first obtained through SCFT calculations within a unit cell\,\cite{cochran2006stability, jiang2010spectral, arora2016broadly,he2025pscf}. Grains with different orientations are then constructed by applying rigid-body rotations $\bR_m \in \mbox{SO}(3)$ ($m=1,2$) to the $\phi_{\mathrm{A},\,\mathrm{bulk}}(\br)$. The segment density of the rotated grain $m$, $\phi_{\mathrm{A},\,\mathrm{grain}}^{\bR_m}(\br)$, is given by,
\begin{equation}\label{grains}
\phi_{\mathrm{A},\,\mathrm{grain}}^{R_m}(\br)=\phi_{\mathrm{A},\,\mathrm{bulk}}\left(\bR_m^T \br\right)=\sum_{\bk \in \mathbb{Z}^{3}}\hat{\phi}_{\mathrm{A},\,\mathrm{bulk}}(\bk) e^{i\left(\bR_m \mathcal{B} \bk\right)^{T} \br},
\end{equation}
where $\bk=(k_1,k_2,k_3)$ is an integer index vector, $\mathcal{B}$ denotes the primitive reciprocal lattice.

In the current study, we focus on the case where the two grains share a common period in the $y$–$z$ plane. 
This case holds if the columns of the matrix $(\tilde \bR_1^T \mathcal B,\, \tilde \bR_2^T \mathcal B)$ can be expressed as rational linear combinations of two linearly independent vectors.
Here, $\tilde \bR_m$ ($m=1,2$) denotes the matrix formed by selecting the last two columns of the rotated matrix $\bR_m$, and $\tilde \bR_m^T \mathcal B$ represents the reciprocal lattice vectors of grain $m$ projected onto the $y$-$z$ plane.

The initial segment density of the A-block in GB structure, $\phi_{\mathrm{A},\,\mathrm{GB}}(\br)$, is generated by a weighted combination of the two rotated grains,
\begin{equation}
    \begin{aligned}
         \phi_{\mathrm{A},\,\mathrm{GB}}(\br)&=(1-c(x))\phi_{\mathrm{A},\,\mathrm{grain}}^{\bR_1}(\br)+c(x)\phi_{\mathrm{A},\,\mathrm{grain}}^{\bR_2}(\br),
    \end{aligned}
\end{equation}
where the smooth function $c(x)$ is given by,
\begin{equation}
    c(x) = \frac{1 + \tanh(\sigma x)}{2},
\end{equation}
and the parameter $\sigma$ is chosen large enough such that $c(-L_x/2)=0$ and $c(L_x/2)=1$.
The GB segment density of the B-block, $\phi_{\mathrm{B},\,\mathrm{GB}}(\br)$, is determined by the incompressibility condition. 
These segment densities provide suitable initial values for SCFT calculations of GB structures.

For solving the MDEs, we adopt periodic boundary conditions in the $x$-direction to approximate the inhomogeneous Dirichlet boundary condition. We will explicitly verify that this approximation does not affect the converged GB structures and the high-precision computation of GB energy.
Consequently, we can treat the spatial variables using the pseudo-spectral method, while the chain contour direction is discretized by the fourth-order backward differentiation method\,\cite{ranjan2008linear,arora2016broadly,he2025pscf}.
Furthermore, we employ the Anderson mixing method to obtain the SCFT solutions corresponding to ordered GB structures\,\cite{thompson2004improved, jiang2015self}. The computational box is kept fixed throughout the SCFT calculations of the GBs.
The entire algorithm is implemented in C++, with parallel computations performed using the FFTW-MPI package\,\cite{frigo2005design, Wang2025SCFT}.
The follwoing set of parameters are chosen based on extensive test. The chain contour is divided into 300 points.
For each GB calculation, the spatial grid is chosen as $N_x \times N_y \times N_z$, where $N_x$, $N_y$, and $N_z$ are taken to be approximately three times the corresponding box lengths $L_x$, $L_y$, and $L_z$, respectively. For example, for the (422) TB calculation with $L_x=145.74R_g,\,L_y=14.72R_g$, and $L_z=24.04R_g$, the spatial grid is $436\times44\times72$. The chosen contour and spatial discretizations ensure a converged free energy with an accuracy of $10^{-6}$, while avoiding high computational cost.

Since periodic boundary conditions are imposed along the $x$-direction, the equilibrium phases obtained from SCFT calculation exhibit distortions near the computational boundaries. 
To accurately investigate the stability of GBs, it is necessary to eliminate the deformation effects introduced by boundary conditions. To this end, we need to identify these deformation regions and exclude them from subsequent analyses. Specifically, we quantify the structural deviation along the $x$-direction by defining the maximum segment density difference $D(x)$ between the GB and the rotated grains, as expressed below, 
\begin{equation}
  D(x) =
\begin{cases}
 \underset{\tilde{\br}}{\max} \left| \phi_{\mathrm{A}, \mathrm{GB}}(x, \tilde{\br}) - \phi_\mathrm{A}^{R_1}(x, \tilde{\br}) \right|, & - L_x/2 \leq x < 0, \\[8pt]
 \underset{\tilde{\br}}{\max} \left| \phi_{\mathrm{A}, \mathrm{GB}}(x, \tilde{\br}) - \phi_\mathrm{A}^{R_2}(x, \tilde{\br}) \right|, &~~~ 0 < x \leq L_x/2,
\end{cases} 
\label{phi_err}
\end{equation}
where $\tilde{\br}=(y,z)^T$. When the value of $D(x)$ satisfies $D(x)\leq \epsilon$ (a prescribed threshold), the local morphology is considered to be consistent with that of the corresponding bulk structure, and the region is identified as an undeformed grain region. Conversely, when $D(x)$ exceeds this threshold, the structure deviates significantly from the bulk morphology and is classified as a deformed region influenced by the GB or the imposed boundary conditions. More detailed calculations are provided in the \textit{GB Energy} subsection.

\section{Results and Discussion} 
\begin{figure}[htbp]
	\centering
	\includegraphics[width=1\linewidth]{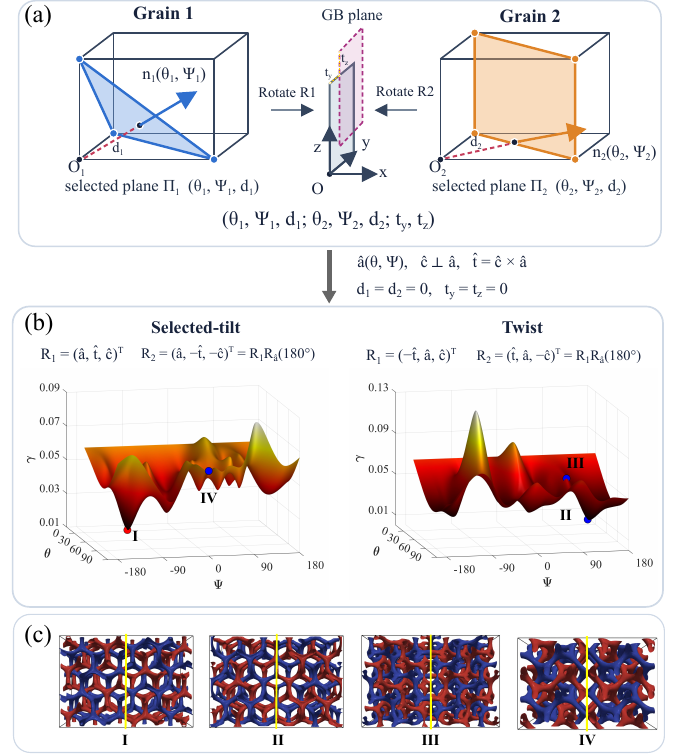}
	\caption{(a) Schematic of the eight-parameter GB construction. (b) GB energy landscapes for prescribed selected-tilt and twist configurations as a function of rotation-axis angles $\theta$ and $\Psi$ at $\chi N=14, f=0.40$.
	(c) Representative GB morphologies corresponding to the labeled in (a), where $[\bar{1}\bar{1}1]$ ATNS (I), $[1\bar{2}\bar{1}]$ TNS (II), $[0\bar{1}\bar{1}]$ TNS (III), and $[100]$ ATNS (IV), with GB energies $\gamma = 0.021,\,0.025,\,0.060$, and $0.061$, respectively. Vertical yellow lines denote the contact planes. The three-dimensional structures are visualized by isosurfaces of the A-block segment density at $\phi_{\mathrm{A},\,\mathrm{GB}} = 0.8$, which clearly depict the network architectures. Blue and red colors are used to distinguish the left- and right-handed networks, respectively.}
	\label{landscape}
\end{figure}

Based on the SCFT framework described above, we investigate the structure and stability of GBs in the DG phase. To search for new and more stable configurations, we construct a GB energy landscape.
Geometrically, a general GB configuration can be constructed by selecting one crystallographic cutoff plane and displacement for each of the two grains. For Grain $i$, the orientation of the cutoff plane is specified by its unit normal $\mathbf n_i(\theta_i,\Psi_i)$, and $d_i$ denotes the signed distance of the plane from the unit-cell origin $O_i$ along its normal direction. After the two selected cutoff planes are brought into coincidence to form a common GB plane, the two grains can still be translated relative to each other within the GB plane, specified by $(t_y,t_z)$. Therefore, a general GB configuration can be parameterized as
\begin{equation}
    (\theta_1,\Psi_1,d_1;\,
     \theta_2,\Psi_2,d_2;\,
     t_y,t_z),
\end{equation}
which defines an eight-dimensional configuration space\,\cite{sutton2015five, sutton1995interfaces}, as illustrated in Figure\,\ref{landscape}a. Exhaustive sampling of this high-dimensional space is computationally impossible. We therefore consider two prescribed two-dimensional subsets of the general GB configuration space.

Specifically, we use the relative orientation between the two grains to parameterize the sampled GB configurations, since varying the relative orientation intuitively changes the relative geometric arrangement of the two DG networks across the GB.
The misorientation angle is fixed at $\omega=180^\circ$, and the direction of the relative-rotation axis $\hat{\mathbf a}$ is taken as the remaining orientational degree of freedom and parameterized by the spherical angles
$(\theta,\Psi)$. The relative orientation of the two grains is then directly
determined by the rotation axis as $R_{\hat{\mathbf a}(\theta,\Psi)}(180^\circ)$. Meanwhile, the cutoff planes are selected to pass through the respective unit-cell origins, giving $d_1=d_2=0$, and the relative in-plane translation is fixed at $t_y=t_z=0$. For each sampled rotation axis $\hat{\mathbf a}(\theta,\Psi)$, an auxiliary crystallographic direction $\hat{\mathbf c}$ perpendicular to $\hat{\mathbf a}$ is determined according to a prescribed rule, and the third direction is defined as $\hat{\mathbf t}=\hat{\mathbf c}\times\hat{\mathbf a}$. Based on these three mutually orthogonal directions, one twist and one selected-tilt GB are constructed for each sampled $(\theta,\Psi)$. For the twist construction, the cutoff plane is perpendicular to the rotation axis $\hat{\mathbf a}$, and the rotation matrices of the two grains are,
\begin{equation}
    R_1^{\mathrm{twist}}
    =
    \left(
        \hat{\mathbf a},
        \hat{\mathbf t},
        \hat{\mathbf c}
    \right)^{\mathrm T},
    \qquad
    R_2^{\mathrm{twist}}
    =
    \left(
        \hat{\mathbf a},
        -\hat{\mathbf t},
        -\hat{\mathbf c}
    \right)^{\mathrm T}
    =
    R_1^{\mathrm{twist}}
    R_{\hat{\mathbf a}}(180^\circ).
\end{equation}
For the selected-tilt construction, multiple cutoff planes can contain the same rotation axis $\hat{\mathbf a}$. We therefore select one representative cutoff plane containing both $\hat{\mathbf a}$ and the prescribed auxiliary direction $\hat{\mathbf c}$. The corresponding rotation matrices are,
\begin{equation}
    R_1^{\mathrm{tilt}}
    =
    \left(
        -\hat{\mathbf t},
        \hat{\mathbf a},
        \hat{\mathbf c}
    \right)^{\mathrm T},
    \qquad
    R_2^{\mathrm{tilt}}
    =
    \left(
        \hat{\mathbf t},
        \hat{\mathbf a},
        -\hat{\mathbf c}
    \right)^{\mathrm T}
    =
    R_1^{\mathrm{tilt}}
    R_{\hat{\mathbf a}}(180^\circ).
\end{equation}
The rotation matrices provide the orientations of the two grains.
Specifically, the first row of each rotation matrix gives the unit normal of the corresponding cutoff plane and aligns it with the laboratory $x$ direction, while the second and third rows give two mutually orthogonal crystallographic directions within the cutoff plane and align them with the laboratory $y$ and $z$ directions, respectively.
The two matrices also satisfy $R_1^{\mathrm T}R_2=R_{\hat{\mathbf a}(\theta,\Psi)}(180^\circ)$ and therefore retain the prescribed relative orientation between the two grains. Together with the constraints $d_1=d_2=0$ and $t_y=t_z=0$, $R_1$ and $R_2$ uniquely define the initial GB configuration.

To construct the GB energy landscapes, 33 rational crystallographic directions are selected as the rotation axes in the parameter space $(\theta,\Psi)\in[0^\circ,90^\circ]\times[-180^\circ,180^\circ)$. The distribution of sampling points is shown in Figure\,S3c. For each sampled point $(\theta,\Psi)_j$, $j=1,\ldots,33$, one twist and one selected-tilt GB are constructed, resulting in a total of 66 independent initial GB configurations. The corresponding rotation matrices $R_1$ and $R_2$ are used to construct the initial fields for the SCFT calculations, after which each GB configuration is independently relaxed to convergence at $\chi N=14, f=0.40$ and its GB energy is calculated via the linear relationship between the excess free energy and the reciprocal of the domain size along the $x$-direction (see \textit{GB Energy} for further details). The 33 sampling points therefore yield two sets of discrete GB energies,
\begin{equation}
    \left\{
        \gamma^{\mathrm{twist}}((\theta,\Psi)_j)
    \right\}_{j=1}^{33},
    \qquad
    \left\{
        \gamma^{\mathrm{tilt}}((\theta,\Psi)_j)
    \right\}_{j=1}^{33}.
\end{equation}
The two sets of discrete energies are then interpolated separately over the $(\theta,\Psi)$ parameter space to obtain the twist and selected-tilt two-dimensional GB energy landscapes shown in Figure\,\ref{landscape}b. Further details of the GB parameterization, the selection of the auxiliary direction $\hat{\mathbf c}$, the construction of the rotation matrices, and the generation of the initial SCFT configurations are provided in Section\,S1 of the Supporting Information. 
Several restrictions on the resulting energy landscapes should be noted. Configurations are excluded when the relative operation between the two grains is equivalent to a bulk symmetry operation or a lattice translation of the DG phase, such that the density fields of the two grains match continuously across $x=0$ without generating an interfacial mismatch. These cases correspond to a single bulk DG morphology rather than a genuine GB. In addition, the GB energy landscapes constructed here are restricted to commensurate GB configurations in which the two grains share a common period in the $y$--$z$ plane. We further emphasize that the twist and selected-tilt families considered here represent two prescribed two-dimensional subsets of the full eight-dimensional GB configuration space rather than all possible DG GBs. In particular, for each rotation axis $\hat{\mathbf a}(\theta,\Psi)$, the selected-tilt construction selects only one representative cutoff plane from the multiple planes containing the same rotation axis and therefore does not include all possible tilt configurations.

The resulting energy landscapes exhibit complex and corrugated behaviour with multiple local minima.
Notably, the tilt and twist energy landscape each reveal a pronounced minimum, labeled I and II, respectively. Minimum I ($\gamma = 0.021$) corresponds to an asymmetric tilt network-switching GB ($[\bar{1}\bar{1}1]$ ATNS), which is energetically more favorable than the experimentally observed $(422)$ TB associated with minimum II ($\gamma = 0.025$), here referred to as a twist network-switching GB ($[1\bar{2}\bar{1}]$ TNS). This finding contrasts with hard-matter systems, where symmetric TBs are typically favored, indicating a different stability mechanism for soft network GBs.

To uncover the physical origin of this stability, we select four representative network-switching GBs from the two energy landscapes for in-depth analysis, including two asymmetric tilt GBs (the low-energy $[\bar{1}\bar{1}1]$ ATNS and the high-energy $[100]$ ATNS) and two twist GBs (the low-energy $[1\bar{2}\bar{1}]$ TNS and the high-energy $[0\bar{1}\bar{1}]$ TNS).
Their three-dimensional morphologies, shown in Figure\,\ref{landscape}c, exhibit network chirality switching across the contact plane. 
The geometric relations between the two grains are summarized in Table\,\ref{tab1}. 
Notably, for the $[1\bar{2}\bar{1}]$ TNS, a space-group symmetry operation was applied so that the resulting crystallographic indices are consistent with those reported experimentally.
Within the thermodynamically stable DG region ($N=100$, $\chi=0.14$, $0.38 \leq f \leq 0.41$), we systematically analyze these GBs from both energetic and structural perspectives, including interaction and entropy contributions, chain trajectories, network topologies, and geometric deformations.

\begin{table}[htbp]
	\caption{Posing of two grains for the four network-switching GBs in DG phase, where $a$ denotes the cubic unit cell size of the bulk DG.}
	\label{tab1}
	\begin{tabular*}{1\textwidth}{@{\extracolsep{\fill}}cccccc}
		\toprule
		& Two & Contact & \multicolumn{2}{c}{Coincident} &  Coincident \\
		& Grains & Plane & \multicolumn{2}{c}{Directions} & Point \\
		\midrule
		\multirow{2}{*}{$[\bar{1}\bar{1}1]$ ATNS} & grain 1 & $(\bar{1}\bar{1}\bar{2})$ & $[\bar{1}\bar{1}1]$ & $[\bar{1}10]$ & $(0,0,0)$ \\
		& grain 2 & $(112)$ & $[\bar{1}\bar{1}1]$ & $[1\bar{1}0]$ & $(0,0,0)$\\
		\midrule
		\multirow{2}{*}{$[1\bar{2}\bar{1}]$ TNS} & grain 1 & $(422)$ & $[1\bar{1}\bar{1}]$ & $[01\bar{1}]$ & $(0,0,a/2)$\\
		& grain 2 & $(422)$ & $[\bar{1}11]$ &$[0\bar{1}1]$& $(0,0,a/2)$ \\
		\midrule
		\multirow{2}{*}{$[0\bar{1}\bar{1}]$ TNS} & grain 1 &  $(0\bar{1}\bar{1})$ & $[0\bar{1}1]$ & $[\bar{1}00]$ & $(0,0,0)$\\
		& grain 2 & $(0\bar{1}\bar{1})$ & $[01\bar{1}]$ & $[100]$ & $(0,0,0)$\\
		\midrule
		\multirow{2}{*}{$[100]$ ATNS} & grain 1 & $(00\bar{1})$ & $[100]$ & $[0\bar{1}0]$ & $(0,0,0)$\\
		& grain 2 & $(001)$ & $[100]$ & $[010]$ & $(0,0,0)$\\
		\bottomrule
	\end{tabular*}
\end{table}

\subsection{GB Energy}\label{sec:energy} 
The relative stability of the four network-switching GBs is quantified by their GB energies.
The GB energy $\gamma$ is defined as the excess free energy of the GB system relative to the rotated grain.
For a region of length $L$ in the $x$-direction, it is given by
\begin{equation}
	\gamma=(E_{\mathrm{GB}}-E_{\mathrm{grain}})L,
	\label{energy_a}
\end{equation}
where $E_{\mathrm{GB}}=F_{\mathrm{GB}}/(nk_{\mathrm{B}}T)$ and
$E_{\mathrm{grain}}=F_{\mathrm{grain}}/(nk_{\mathrm{B}}T)$ are the free energies per chain of the GB system and the corresponding rotated grain, respectively.
%the dimensionless free energies per chain, or equivalently the free-energy densities nondimensionalized by $\rho_0 k_{\mathrm{B}}T/N$.

This relationship implies that the free energy difference scales linearly with the inverse of length,
\begin{equation}
	E_{\mathrm{GB}}-E_{\mathrm{grain}}=\gamma L^{-1},
	\label{energy_b}
\end{equation}
where the slope of the linear fit yields the GB energy $\gamma$.

\begin{figure}[htbp]
	\centering
	\includegraphics[width=1\linewidth]{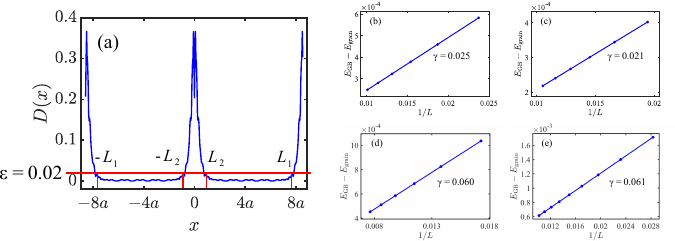}
 	\caption{(a) Maximum density difference between the $(422)$ TNS and the corresponding grains along $x$-direction, where $|x| < L_2$ is the the central GB region, $-L_1 \leq x \leq -L_2$ and $L_2 \leq x \leq L_1$ correspond to grain 1 and grain 2, respectively, while $|x| > L_1$ are distortion regions resulting from the boundary conditions. The optimal edge length of the bulk DG is $a = 8.5$ at $f = 0.4$\,\cite{matsen2012effect}. Free energy differences of (b) $[1\bar{2}\bar{1}]$ TNS, (c) $[\bar{1}\bar{1}1]$ ATNS, (d) $[0\bar{1}\bar{1}]$ TNS and (e) $[100]$ ATNS relative to their rotated grain 1 as a function of $1/L$, where the slope of the linear fit is the GB energy $\gamma$.}
	\label{energy_calculate}
\end{figure}

To ensure an accurate extraction of the GB energy, it is crucial to eliminate boundary-induced distortions arising from the imposed boundary conditions. Taking the $[1\bar{2}\bar{1}]$ TNS at $f = 0.4$ and $\chi N=14$ as an example, we employ eq.\,\ref{phi_err} with a threshold of $\epsilon=0.02$ to identify the deformed regions. As shown in Figure\,\ref{energy_calculate}a, regions with $|x| > L_1$ are influenced by boundary constraints, while the region $|x| < L_2$ corresponds to the central GB region. Therefore, only the intermediate domain $|x| \leq L_1$ is retained for the GB energy calculation. 
Within the window $2L_2 < L < 2L_1$, we evaluate the free energy $E_{\mathrm{GB}}$ for a series of lengths $L$ by restricting the converged SCFT solutions $\phi_{\alpha, \mathrm{GB}}(\br), w_{\alpha}(\br), Q$, $\alpha\in\{ \mathrm{A}, \mathrm{B}\}$ to the interval $-L<x<L$ and substituting them into the free energy functional defined in eq.\,\ref{free_energy}. 
The corresponding grain free energy $E_{\mathrm{grain}}$ is obtained following the same procedure.
The resulting free energy difference $E_{\mathrm{GB}} - E_{\mathrm{grain}}$ is plotted as a function of $L^{-1}$ in Figure\,\ref{energy_calculate}b. The excellent linear scaling confirms that residual boundary effects are effectively eliminated, enabling an accurate determination of the GB energy.
For the $[1\bar{2}\bar{1}]$ TNS, the GB energy is extracted as $\gamma = 0.025$ from the slope of the linear fit (Figure\,\ref{energy_calculate}b). Under the same parameters, the $[\bar{1}\bar{1}1]$ ATNS, $[0\bar{1}\bar{1}]$ TNS and $[100]$ ATNS exhibit analogous scaling behavior, but their GB energies differ markedly. The $[\bar{1}\bar{1}1]$ ATNS has a lower value of $\gamma = 0.021$ (Figure\,\ref{energy_calculate}c), while the $[0\bar{1}\bar{1}]$ TNS and $[100]$ ATNS exhibit larger GB energies of $\gamma = 0.060$ and $0.061$, respectively (Figures\,\ref{energy_calculate}d-e). All calculated GB energies converge within an accuracy of $10^{-4}$, which is sufficient to determine the relative thermodynamic stability of these configurations.

\begin{figure}[htbp]
	\centering
	\includegraphics[width=1\linewidth]{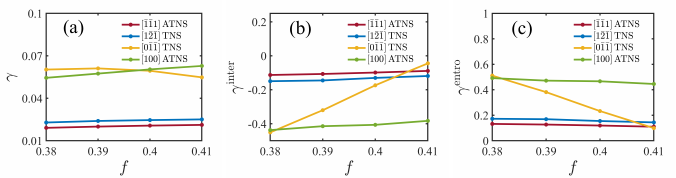}
	\caption{(a) GB energy $\gamma$, (b) interaction energy $\gamma^{\mathrm{inter}}$, and (c) entropy energy $\gamma^{\mathrm{entro}}$ as a function of the volume fraction $f$.}
	\label{energy_f}
\end{figure}

Using the scaling method, the GB energies of all four structures as a function of the A-block volume fraction $f$ are plotted in Figure\,\ref{energy_f}a. Across the composition range $0.38 \leq f \leq 0.41$, the $[\bar{1}\bar{1}1]$ ATNS consistently possesses the lowest GB energy, demonstrating its superior thermodynamic stability. The $[1\bar{2}\bar{1}]$ TNS exhibits slightly higher and comparably weak $f$-dependent GB energies, whereas the $[0\bar{1}\bar{1}]$ TNS and $[100]$ ATNS are significantly less stable and exhibit greater sensitivity to composition.
To elucidate the physical origin of these stability differences, we decompose the total GB energy into interaction energy ($\gamma^{\mathrm{inter}}$) and entropy energy ($\gamma^{\mathrm{entro}}$) contributions, 
\begin{equation}
    \begin{aligned}
    \gamma^{\mathrm{inter}} &= L\left(F^{\mathrm{inter}}_{\mathrm{GB}}- F^{\mathrm{inter}}_{\mathrm{grain}}\right)/ nk_{\mathrm{B}}T,
    \\
    \gamma^{\mathrm{entro}} &= 
    L\left[\left(-TS_{\mathrm{GB}}\right)-\left(- TS_{\mathrm{grain}}\right)\right] / nk_{\mathrm{B}}T,
    \label{energy_gb_two_part}
    \end{aligned}
\end{equation}
where $F^{\mathrm{inter}}_{\mathrm{GB}}/nk_{\mathrm{B}}T$ ($F^{\mathrm{inter}}_{\mathrm{grain}}/nk_{\mathrm{B}}T$) and $-TS_{\mathrm{GB}}/nk_{\mathrm{B}}T$ ($-TS_{\mathrm{grain}}/nk_{\mathrm{B}}T$) represent the interaction and entropy energies of GB structure (grain 1), respectively.

As shown in Figures\,\ref{energy_f}b-c, the $[0\bar{1}\bar{1}]$ TNS and $[100]$ ATNS exhibit much stronger composition dependence in both interaction and entropy energies than the $[\bar{1}\bar{1}1]$ ATNS and $[1\bar{2}\bar{1}]$ TNS, explaining their greater sensitivity of the total GB energies to $f$. Furthermore, the interaction energies of the $[\bar{1}\bar{1}1]$ ATNS and $[1\bar{2}\bar{1}]$ TNS are comparable, with the $[\bar{1}\bar{1}1]$ ATNS even exhibiting a slightly higher interaction penalty. However, this unfavorable interaction contribution is more than offset by a substantially lower entropy penalty, which minimizes the total free energy of the $[\bar{1}\bar{1}1]$ ATNS. In contrast, although the $[100]$ TNS and $[011]$ ATNS gain partial energetic advantage from reduced interaction energies, their large entropy penalties dominate, resulting in higher total GB energies.
These results demonstrate that, the GB stability in BCP network phases is governed primarily by entropy rather than interaction energy, in sharp contrast to hard matter.

\subsection{Chain Trajectories}
\label{chain}
\begin{figure}[htbp]
	\centering
	\includegraphics[width=1\linewidth]{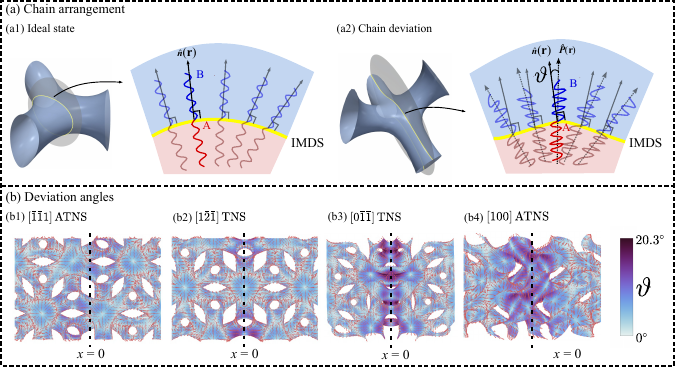} 
	\caption{(a) Schematics of chain trajectory packing and the definition of deviation angle $\vartheta$.
	(a1) and (a2) depict the chain arrangements in the ideal state and the chain deviation near new nodes, respectively.
	The deviation angle $\vartheta$ is defined at the isosurface $\phi_{\mathrm{A}}(\br)=0.5$ as the angle between the normalized polar order parameter $\hat{\bP}(\br)$ and the local surface normal $\hat{\bn}(\br)$.
	(b) Spatial distributions of the deviation angle $\vartheta$ in the (b1) $[\bar{1}\bar{1}1]$ ATNS, (b2) $[1\bar{2}\bar{1}]$ TNS, (b3) $[0\bar{1}\bar{1}]$ TNS and (b4) $[100]$ ATNS.
	The red arrows represent the deviation vectors projected onto the $\phi_\mathrm{A}(\br) = 0.5$ surface, where the arrow length corresponds to the magnitude of the local chain deviation.}
	\label{angle_3d}
\end{figure}

The dominance of entropy in determining GB stability motivates a microscopic analysis of chain conformations and network geometry near the interface. In ideal BCP networks, polymer chains preferentially orient along the normal direction of intermaterial dividing surface (IMDS, the isosurface $\phi_\mathrm{A}(\br) = 0.5$)\,\cite{greenvall2023chain, dimitriyev2023medial}. This arrangement represents a relaxed conformational state with minimal packing frustration (see Figure\,\ref{angle_3d}a). However, in the network-switching GBs, the geometric mismatch between the two grains is alleviated through conformational adjustment of flexible chains, leading to deviations of chain orientation from the IMDS normal direction and an associated entropy penalty (see Figure\,\ref{angle_3d}a).
Meanwhile, the adjustment of chain conformations is accompanied by geometric deformation of the soft network. This alters the contact patterns and distribution of A–B monomers, affecting the interaction energy of the system. 

To quantify these effects, we introduce a polar order parameter to describe the average orientation of polymer chains\,\cite{greenvall2023chain, dimitriyev2023medial},
\begin{equation}
\bP(\br)=\int_{0}^{1}\bJ(\br,s)\mathrm{d}s,
\end{equation}
where the chain flux $\bJ(\br,s)$ is given by,
\begin{equation}
\bJ(\br,s)=\frac{1}{6Q}(q(\br,s)\nabla q^{\dagger}(\br,1-s)-q^{\dagger}(\br,1-s)\nabla q(\br,s)),\,s \in [0, 1].
\end{equation}
The chain conformational frustration is characterized by the chain deviation angle $\vartheta$ between the normalized polar order parameter $\hat{\bP}(\br)$ and the unit normal vector of the IMDS $\hat{\bn}(\br)$ (see Figure\,\ref{angle_3d}a). Meanwhile, variations in the interaction energy are quantified by the IMDS area per unit volume $A/V^L$, where $V^L$ is the volume of the measurement domain. Figure\,\ref{angle_3d}b visualizes the spatial distribution of these deviation angles for all four GBs. Large deviations are localized near the contact plane, where network reconnection and topological rearrangement occur.

\begin{figure}[htbp]            
	\centering
	\includegraphics[width=0.9\linewidth]{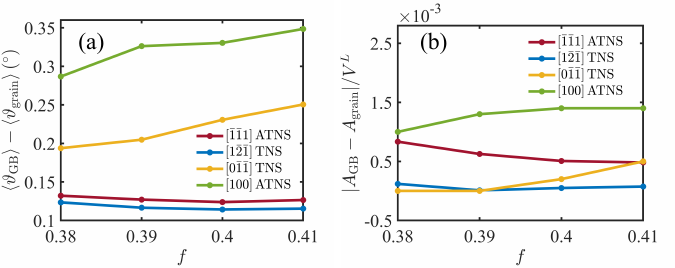}
	\caption{(a) Average chain deviation angle and (b) IMDS area of the GBs with respect to grain 1, as a function of the volume fraction $f$.}
	\label{chain_angle}
\end{figure}

As shown in Figure\,\ref{chain_angle}, the average chain deviation angle and the IMDS area qualitatively characterize the extent of chain conformational rearrangement and IMDS geometric deformation, respectively. Compared with the low-energy $[\bar{1}\bar{1}1]$ ATNS and $[1\bar{2}\bar{1}]$ TNS, the high-energy $[0\bar{1}\bar{1}]$ TNS and $[100]$ ATNS exhibit larger variations in both quantities over the composition range, indicating more pronounced chain conformational rearrangement and IMDS geometric deformation with increasing $f$. Larger chain conformational rearrangements can influence the conformational entropy by modifying the configurational freedom of polymer chains, whereas greater IMDS geometric deformation can influence the monomer interaction energy by altering the contact pattern between A and B monomers. Consequently, both the entropy and interaction energy exhibit stronger composition dependence in the high-energy GBs. In contrast, the $[\bar{1}\bar{1}1]$ ATNS and $[1\bar{2}\bar{1}]$ TNS maintain relatively small chain deviation angles and weak variations in the IMDS area, consistent with their much weaker composition dependence in both entropy and interaction energies.

\begin{figure}[htbp]
	\centering
	\includegraphics[width=1\linewidth]{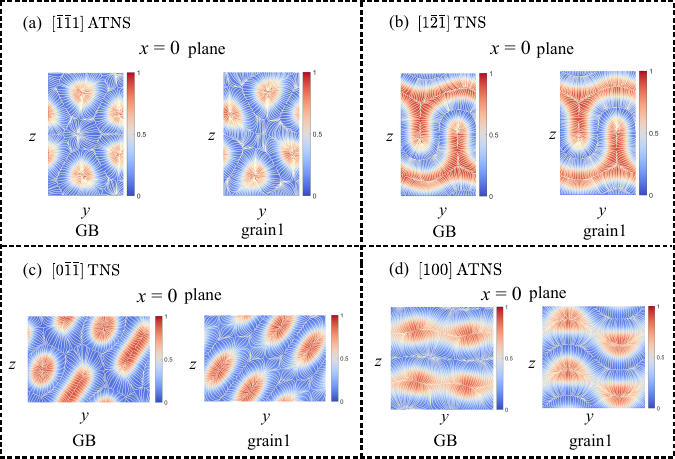}
	\caption{Sections $x=0$ of chain trajectories for four GBs (left) and the corresponding grain 1 (right) at $f = 0.40,\,\chi N=14$. The color bar indicates the density distribution of component A.}
	\label{angle_2d}
\end{figure}

The visualization of chain trajectories further supports this mechanism. In the $[\bar{1}\bar{1}1]$ ATNS and $[1\bar{2}\bar{1}]$ TNS, chain aggregation patterns near the interface remain highly consistent with those of the adjoining grains (Figures\,\ref{angle_2d}a-b), indicating that only minor conformational adjustments are required to accommodate the orientational mismatch. In contrast, the $[0\bar{1}\bar{1}]$ TNS and $[100]$ ATNS require substantial reconstruction of both chain trajectories and A-rich domain morphologies (Figures\,\ref{angle_2d}c-d). Specifically, the chain trajectories in the $[0\bar{1}\bar{1}]$ TNS undergo a transition from the strip-like aggregation of grain 1 to point-like aggregation (Figure\,\ref{angle_2d}c), while the $[100]$ ATNS transitions from a double short-strip aggregation into a single long-strip aggregation (Figure\,\ref{angle_2d}d). These pronounced rearrangements provide a microscopic explanation for the stronger composition dependence of the entropy and interaction energies in the high-energy GBs. The origin of their substantially larger entropy penalties is closely related to the evolution of the network topology, which is analyzed in the following section.

\subsection{GB Network Topologies}
\label{sec:real-space}
Based on chain-trajectory analysis, we elucidate the microscopic origins of the energetic hierarchy among DG network-switching GBs. 
Unlike hard matter, where lattice mismatch and elastic distortion dominate GB energies, soft network phases possess an additional degree of freedom, {\it i.e.} network-topology reconstruction. This unique feature enables stress relaxation through geometric deformation and topological rearrangements.
A notable feature common to all four GBs studied here is the switching of network handedness near the GB region, reflecting the intrinsic topological flexibility of the DG network.
Therefore, we proceed to a comparative analysis of these GB structures from a topological perspective.
All results in this section are obtained at fixed SCFT parameters $f=0.40$ and $\chi N=14$. 
For all mirror-symmetric GBs discussed below, including $[211]$ and $[0\bar{1}\bar{1}]$ TNSs, their mirror symmetries are verified by ensuring that the density difference satisfies $|\phi_{\mathrm{A},\,\mathrm{GB}}(x,y,z)-\phi_{\mathrm{A},\,\mathrm{GB}}(-x,y,z)|<10^{-13}$.
Here, nodes are identified as local maxima of the converged A-block density field, $\phi_{\mathrm{A}}(\mathbf r)$, whose values are not less than $99\%$ of maximum bulk value.

\subsubsection{$[\bar{1}\bar{1}1]$ asymmetric tilt network-switching (ATNS) GB}
\label{sec:netsw}
\begin{figure}[htbp]
	\centering
	\includegraphics[width=1\linewidth]{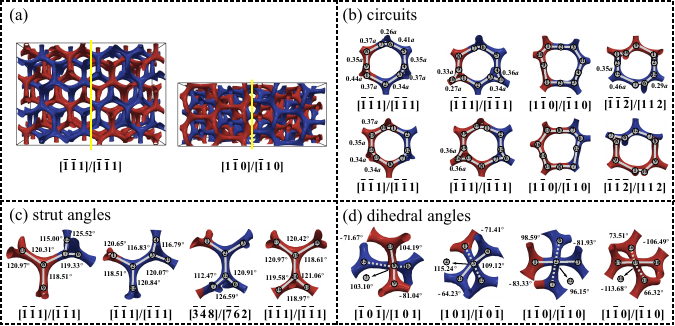}
	\caption{(a) $[\bar{1}\bar{1}1]$ ATNS skeletons. The viewing direction with respect to the left/right grain is labeled below each figure. (b) Eight closed circuits. (c) Strut angles defined by two adjacent struts. (d) Dihedral angle between two planes defined by three consecutive struts connecting four adjacent nodes. Each local configuration is projected along the shared strut of the two planes. Solid line denotes the plane spanned by the shared strut and the forward strut, whereas dashed line denotes the plane spanned by the shared strut and the backward strut. The coreesponding dihedral angle between the two planes are labeled, with ``+'' and ``-'' represent clockwise and counterclockwise, respectively.}
	\label{netsw}
\end{figure}

Among the four investigated DG network-switching GBs, the $[\bar{1}\bar{1}1]$ ATNS exhibits the highest thermodynamic stability. Its network topologies are illustrated in Figure\,\ref{netsw}.
Due to the $180^{\circ}$ rotational symmetry of the two grains about the $[\bar{1}\bar{1}1]$ axis, it is sufficient to analyze one network branch. 
Nodes are labeled by numbers, where identical labels assigned to equivalent positions across periods.
At the contact plane ($x = 0$), the networks of opposing chirality reconnect directly through existing struts, without the formation of new nodes.
This connection produces eight circuits that traverse the $x=0$ plane, containing nine, ten, and eleven nodes, respectively (Figure\,\ref{netsw}b).

Although the chirality reversal induces some distortions in the strut lengths and angles near the interface, these deformations remain spatially localized.
Specifically, struts crossing the $x=0$ plane (e.g., 1-2, 7-8) deviate less than $6\%$ from the bulk length ($0.35a$), whereas those near $x=0$ (e.g., 6-7, 1-9, 1-10, 7-12) undergo larger deformations.
The strut angles near $x=0$ deviate by less than $7\%$ from the equilibrium bulk value of $120^{\circ}$ (Figure\,\ref{netsw}c). The dihedral angles remain close to their complementary bulk reference values of $70.5^{\circ}$ or $109.5^{\circ}$, with the largest deviation observed for 3-2-12-7 in Figure\,\ref{netsw}d.
In the bulk DG, each node lies in a plane named by its three neighboring nodes.
This node coplanarity is well-preserved in the ATNS, with deviations remaining below $4\%$ even near $x=0$ (Figure\,\ref{netsw}d).
The sums of multiple sets of adjacent dihedral angles are close to $180^{\circ}$ (e.g., dihedral angles 1-(2,3)-12, 1-(2,12)-3, and 17-(15,14)-10).
Such preservation of coplanarity and strut-mediated connection allow polymer chains to accommodate network switching through smooth redistribution of contour length and local torsion, rather than through severe bending or orientational confinement.
As a result, the $[\bar{1}\bar{1}1]$ ATNS accommodates network reconnection with the minimal chain packing frustration, leading to a small entropy penalty and explaining its superior thermodynamic stability.

\subsubsection{$[1\bar{2}\bar{1}]$ twist network-switching (TNS) GB}
\label{sec:tb}
\begin{figure}[htbp]
	\centering
    \includegraphics[width=1.0\linewidth]{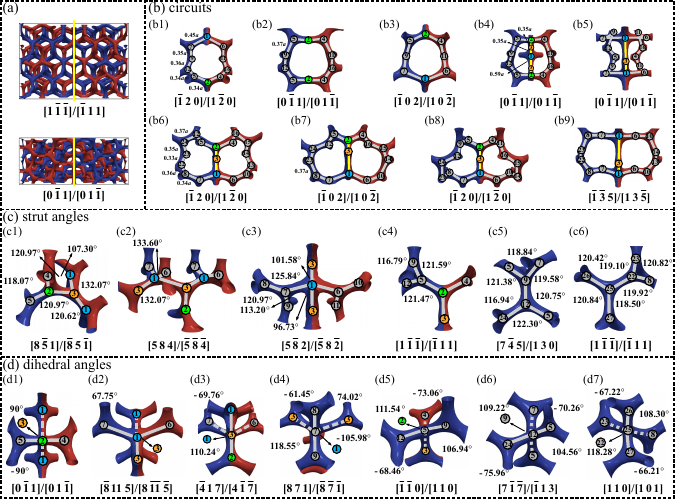}
	\caption{(a) $[1\bar{2}\bar{1}]$ TNS skeletons.
	(b)-(d) Circuits, strut angles, and dihedral angles.}
	\label{tb}
\end{figure}

The $[1\bar{2}\bar{1}]$ TNS is slightly higher in GB energy than the $[\bar{1}\bar{1}1]$ ATNS and exhibits a distinct reconnection mechanism, as shown in Figure\,\ref{tb}.
At the contact plane, reconnection between the two grains generates three types of new nodes, including a four-connected node (labeled blue 1) and two three-connected nodes (labeled green 2 and orange 3).
Node 1 is formed by inward displacement and fusion of two out-of-plane bulk DG nodes. In contrast, nodes 2 and 3 arise from the fusion of in-plane bulk DG nodes from the two opposing chiral networks and remain at the original in-plane node positions.
The circuits associated with these new nodes are classified into two categories, single-circuit and double-circuit. The former arise from direct connections between grains through the new nodes (Figure\,\ref{tb}b), while the latter occur when the circuits from both grains share the new nodes (Figure\,\ref{tb}b).  
Due to the mirror symmetry, only half of the strut lengths are labeled in each circuit. 
These node-connectivity patterns and closed circuits agree with previous studies\,\cite{feng2021visualizing, chen2026computational}, indicating that the computed $[1\bar{2}\bar{1}]$ TNS captures the same local network topology.

The newly formed nodes lead to pronounced geometric reconstruction in their vicinity (Figures\,\ref{tb}b-d).
Most struts deviate by less than $6\%$ from the bulk length, except for struts 1-3 and 1-7 connected to the four-connected Node 1, which show substantial deviation (Figure\,\ref{tb}b).
For instance, the notable elongation of strut 1-3 to $0.59a$ results from the inward offset and fusion of two struts from two grains.
This value is close to the largest boundary-strut length reported experimentally for the reconstructed TB loops, approximately $0.58a$\,\cite{feng2021visualizing}. Relative to the bulk DG strut length of $0.35a$, this corresponds to comparable elongations of approximately $69\%$ in the SCFT calculation and $66\%$ in the experiment. The shortened struts differ more noticeably. The shortest corresponding SCFT struts are about $0.33a$ (strut 22-23 in Figure\,\ref{tb}b), only $5.7\%$ below the bulk value, whereas the experimental distribution extends to about $0.2a$, corresponding to a $43\%$ reduction from the bulk length. These comparisons indicate that the SCFT-relaxed $[1\bar{2}\bar{1}]$ TNS captures the experimentally observed large strut elongation, while the extent of strut shortening is smaller than in the experimental reconstruction.
Similarly, strut angles involving Node 1 deviate substantially, exceeding $10\%$ from the ideal $120^{\circ}$ (Figure\,\ref{tb}c).
Significant deviations of dihedral angles occur when new nodes are involved, and vertical planes appear (Figure\,\ref{tb}d).
Relative to the corresponding experimental mean values, the SCFT dihedral angles in Figure\,\ref{tb}d differ by about $9\%$--$12\%$. More importantly, the values of dihedral angles remain close to the ideal DG value of $70.5^{\circ}$, deviating by only $0.74^{\circ}$--$2.77^{\circ}$, compared with $4.55^{\circ}$--$10.0^{\circ}$ in the experiment of Feng et al.\,\cite{feng2021visualizing}.
The network chirality reverses locally, as indicated by dihedral angles (Figure\,\ref{tb}d). 
The node coplanarity is preserved only in partial regions.
Specifically, when only one new node on $x=0$ is involved, coplanarity can be well maintained (see dihedral angles 8-(7,1)-9, 9-(5,2)-12 in Figure\,\ref{tb}d), while regions without new nodes also display small deviations below $4\%$ of $180^{\circ}$ (see dihedral angles 5-(12,9)-24 and 8-(22-25)-23 in Figure\,\ref{tb}d). 
Compared with the $[\bar{1}\bar{1}1]$ ATNS, the $[1\bar{2}\bar{1}]$ TNS incurs an additional entropic cost associated with new-node generation and partial breakdown of coplanarity, accounting for its higher GB energy despite its well-known stability in hard matter.

\subsubsection{$[0\bar{1}\bar{1}]$ twist network-switching (TNS) GB}\label{sec:twist1}
\begin{figure}[htbp]
	\centering
	\includegraphics[width=1.0\linewidth]{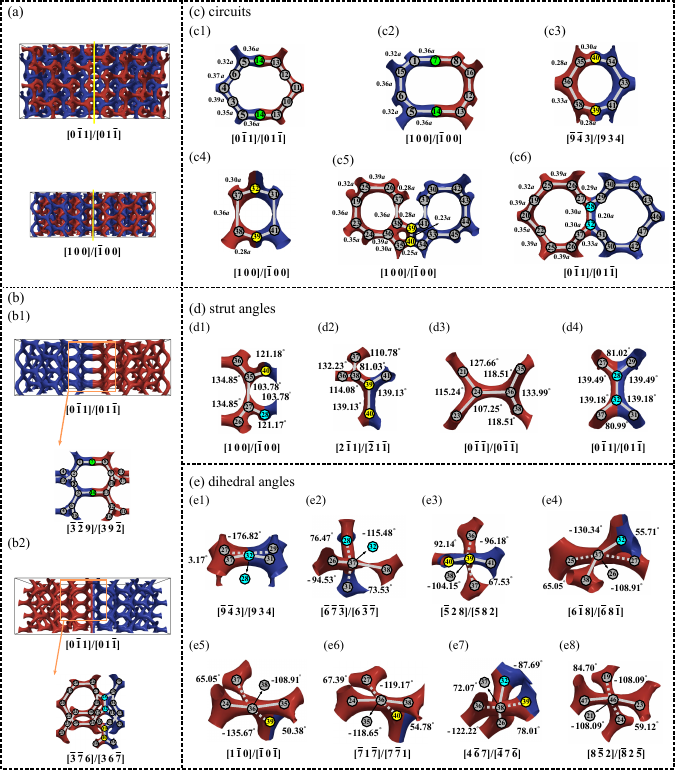}
	\caption{(a) $[0\bar{1}\bar{1}]$ TNS skeletons. (b) Two networks. (c)-(e) Circuits, strut angles, and dihedral angles.}
	\label{twist1}
\end{figure}

We next examine the $[0\bar{1}\bar{1}]$ TNS, a high-energy configuration characterized by extensive topological reconstruction at the GB region.
In contrast to the $[1\bar{2}\bar{1}]$ TNS, the chirality switching in the $[0\bar{1}\bar{1}]$ TNS induces complex node reconstruction and produces three types of new nodes. As shown in Figure\,\ref{twist1}b, the two-connected green nodes are formed by direct fusion of two bulk nodes from the contacting grains, while the three-connected yellow and cyan nodes arise from fusion of two bulk nodes followed by splitting.
These reconstructed nodes and their connecting struts form two categories of circuits (Figure\,\ref{twist1}c), single-circuits c1–c4 and double-circuits c5–c6.
The geometric distortions associated with this reconstructions become severe.
Strut lengths connected to new nodes exhibit large contractions, with strut angles and dihedral angles deviating markedly from their bulk values.
Most importantly, the node coplanarity is extensively destroyed, not only at the contact plane but also in regions far from the interface.

Specifically, struts involving two new nodes shorten by over $34\%$ from $0.35a$ (see struts 39-40 and 28-32 in Figure\,\ref{twist1}c), while those containing a single new node typically shorten by over $14\%$, except for struts 1-7 and 5-14 (Figure\,\ref{twist1}c).
Significant structural distortions occur in network b2 (Figure\,\ref{twist1}b), with deviations in network b1 remaining below $9\%$ and $8\%$ for strut angles and coplanarity, respectively.
Consequently, we only show the angular deformations in network (b2) as illustrated in Figures\,\ref{twist1}d-e.
Strut angles (Figure\,\ref{twist1}d) and dihedral angles (Figure\,\ref{twist1}e) involving new nodes exhibit substantial deformations, while dihedral angles deviate more than strut angles.
If the new nodes are considered as vertices, the deformations of most strut angles exceed $16\%$ (Figure\,\ref{twist1}d).
When nodes connected to $x=0$ (nodes 27 and 38) are involved, geometric parameters change to a greater extent than in the $[1\bar{2}\bar{1}]$ TNS.
For example, the strut angles 37-38-36, 35-36-38 in Figure\,\ref{twist1}d and dihedral angles 38-36-35-27, 36-38-37-26 in Figure\,\ref{twist1}e occur large deviations. 
Even far away from $x=0$, some dihedral angles still vary clearly (an example is angle 47-46-21-19 in Figure\,\ref{twist1}e).
Furthermore, the coplanarity is also largely broken, as can be seen in dihedral angles 37-(32,28)-31, 32-(37,26)-38, and 39-(38,36)-37 (Figure\,\ref{twist1}e). The fusion and splitting of new nodes, and the widespread breakdown of coplanarity, impose strong orientational and bending constraints on flexible polymer chains. This leads to intense chain packing frustration and a large entropy penalty.
Consequently, the $[0\bar{1}\bar{1}]$ TNS exhibits a higher GB energy than both the ATNS and the $[1\bar{2}\bar{1}]$ TNS.

\subsubsection{$[100]$ asymmetric tilt network-switching (ATNS) GB}\label{sec:at2}

\begin{figure}[htbp]
    \centering
    \includegraphics[width=1\linewidth]{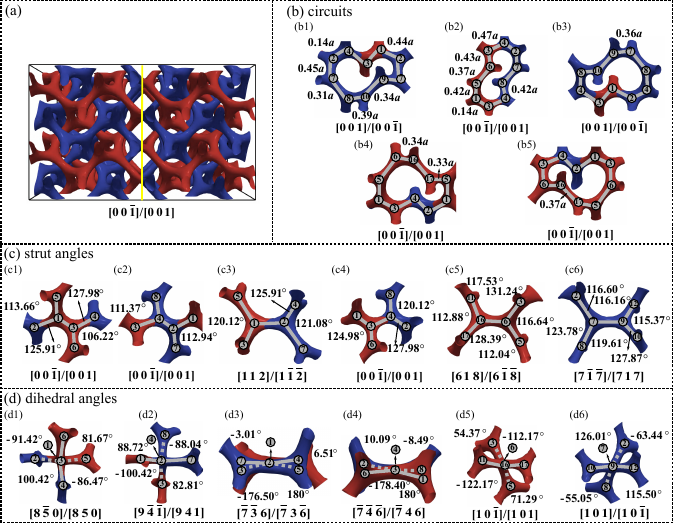}
    \caption{(a) The $[100]$ ATNS skeleton. (b)-(d) Circuits, strut angles, and dihedral angles.}
    \label{atns2}
\end{figure}

Finally, we analyze the $[100]$ ATNS, which has the highest GB energy among the four configurations. 
As illustrated in Figure\,\ref{atns2}, similar to the $[\bar{1}\bar{1}1]$ ATNS, its topological reconstruction is achieved through the reconnection of existing struts without the formation of new nodes. However, the associated geometric distortions are more pronounced. The adjacent bulk nodes and the struts connecting them form five closed circuits that cross the $x=0$ plane.

Unlike $[\bar{1}\bar{1}1]$ ATNS, the chirality inversion in $[100]$ ATNS induces severe distortions in the strut lengths and angles near the contact plane, with appreciable deviations persisting even into the more distant bulk regions. Notably, struts crossing the $x=0$ plane (struts 1-2 and 3-4) are elongated by more than $20\%$, whereas the neighboring struts (struts 2-4 and 1-3) contract by over $50\%$ to alleviate the interfacial geometric mismatch (Figure\,\ref{atns2}b). Such deviations are also evident in the angular data, where dihedral angles (Figure\,\ref{atns2}d) and strut angles (Figure\,\ref{atns2}c) associated with the struts passing through the $x=0$ plane exhibit appreciable deviations from their bulk values. The deviations are particularly pronounced for the dihedral angles, among which angles 4-3-1-2, 6-3-1-2, 8-2-4-1 in Figure\,\ref{atns2}d.
Furthermore, the chirality of the network is locally reversed, as indicated by the dihedral angles in Figure\,\ref{atns2}d. 

More importantly, the deviation from node coplanarity in the $[100]$ ATNS extends over a broader region than in the other three GB configurations. The deviations are observed not only for local instances involving struts that pass through the $x=0$ plane, such as dihedral angles 6-(3,1)-4 and 8-(4,2)-3 in Figure\,\ref{atns2}d, but also for cases that do not involve nodes directly connected across the $x=0$ plane, such as 11-(16,6)-17 and 10-(9,7)-12 in Figure\,\ref{atns2}d.
In other words, the loss of coplanarity persists into the adjacent bulk region, where it is more pronounced than in the corresponding regions of the other three GBs. These broadly distributed geometric distortions indicate that polymer chains in the $[100]$ ATNS are subjected to conformational constraints over a much larger spatial region, resulting in more severe chain-packing frustration and a larger entropy penalty than in the other three GBs.

\subsection{Geometric Deformation Statistics}
\label{sec:shape}
To quantitatively correlate geometric deformation with GB stability, we analyze the maximum geometric deviations in strut lengths, strut angles, dihedral angles, and node coplanarity for the four DG GBs (Table\,\ref{tab2}). 
The statistical data spans two grain periods along $x$-direction (centered at $x=0$) and one period along both $y$ and $z$ directions. We evaluate node coplanarity by measuring the maximum deviation of the dihedral angles formed by three planes, each defined by the node and its two connecting nodes. 
We then partition the nodes into three distinct spatial regions.
Region I corresponds to nodes on $x=0$, region N to nodes with a strut touching or crossing $x=0$, and region F to the remaining nodes.
This classification applies to strut lengths and angles, with elements assigned to region I if constituent node lies there, to region N if node lies in N (but not I), and to region F otherwise.
In the statistical analysis of dihedral angle deviations, we consider only the geometric deformation of struts pre-existing in the original bulk grain structures, excluding metrics associated with newly generated struts.

\begin{table}[htbp]
	\footnotesize
	\caption{Maximum variations for strut lengths, dihedral angles, strut angles, coplanarity.}
	\label{tab2}
	\begin{tabular*}{1\textwidth}{@{\extracolsep{\fill}}cccccc}
		\toprule
		\multirow{2}{*}{GBs} & \multirow{2}{*}{Region} & Strut length & Strut angle  &  Dihedral angle & Coplanarity \\
		&  & $(0.35a)$ & $(120^{\circ})$  &  $(\pm70.5^{\circ})$ & $(180^{\circ})$\\
		\midrule
		\multirow{2}{*}{$[\bar{1}\bar{1}1]$ ATNS} & N & $\pm 0.11a$ & $\pm8.15 ^{\circ}$ & $\pm13.35 ^{\circ}$ & $\pm6.65 ^{\circ}$\\
		& F & $\pm 0.06a$ & $\pm7.69 ^{\circ}$ & $\pm9.76 ^{\circ}$ & $\pm7.19^{\circ}$\\
		\midrule
		\multirow{3}{*}{$[1\bar{2}\bar{1}]$ TNS} & I & $\pm0.24 a$ & $\pm 23.27^{\circ}$ & $\pm35.48 ^{\circ}$ & $\pm69.76 ^{\circ}$\\
		& N & $\pm0.02 a$ & $\pm3.58 ^{\circ}$ & $\pm4.94 ^{\circ}$ & $\pm5.18 ^{\circ}$\\
		& F & $\pm 0.02a$ & $\pm6.57 ^{\circ}$ & $\pm8.78 ^{\circ}$ & $\pm6.32 ^{\circ}$\\
		\midrule
		\multirow{3}{*}{$[0\bar{1}\bar{1}]$ TNS} & I & $\pm0.15 a$ & $\pm16.22 ^{\circ}$ & $\pm26.17 ^{\circ}$ & $\pm20.72 ^{\circ}$\\
		& N & $\pm0.07 a$ & $\pm14.85 ^{\circ}$ & $\pm12.72 ^{\circ}$ & $\pm8.33 ^{\circ}$\\
		& F & $\pm0.04 a$ & $\pm12.75 ^{\circ}$ & $\pm14.21 ^{\circ}$ & $\pm12.79 ^{\circ}$\\
		\midrule
		\multirow{2}{*}{$[100]$ ATNS} & N & $\pm 0.21a$ & $\pm13.78 ^{\circ}$ & $\pm29.92 ^{\circ}$ & $\pm11.85 ^{\circ}$\\
		& F & $\pm 0.04a$ & $\pm9.71 ^{\circ}$ & $\pm15.45 ^{\circ}$ & $\pm13.47^{\circ}$\\
		\bottomrule
	\end{tabular*}
\end{table}

The statistical results reveal a clear distinction between low- and high-energy GBs. For the two twist-type GBs, $[1\bar{2}\bar{1}]$ TNS and $[0\bar{1}\bar{1}]$ TNS, there exists an I-region at the GB center owing to new-node generation and local network-topology reconstruction. In contrast, the two asymmetric tilt GBs, $[\bar{1}\bar{1}1]$ ATNS and $[100]$ ATNS, contain no I-region, and their geometric distortions are distributed only within the N- and F-regions. The formation of new nodes induces local topological reconstruction and forces nearby polymer chains to undergo conformational rearrangement, thereby introducing additional chain-packing frustration and entropy penalties. However, Table\,\ref{tab2} indicates that new-node formation alone does not determine GB stability.

A comparison of the geometric parameters further shows that, for the low-energy $[1\bar{2}\bar{1}]$ TNS, the pronounced geometric distortions are largely confined to the I-region. For example, the coplanarity deviation reaches $69.76^\circ$ in the I-region, whereas it is only $5.18^\circ$ and $6.32^\circ$ in the N- and F-regions, respectively. Similar variation is also observed for the strut length, strut angle, and dihedral angle, with the largest geometric distortions concentrated in the I-region and only minor deviations in the surrounding N- and F-regions. This indicates that the topological reconstruction is localized near the GB center while the geometry of the surrounding network remains largely preserved. Likewise, the low-energy $[\bar{1}\bar{1}1]$ ATNS exhibits only small coplanarity deviations of $6.65^\circ$ and $7.19^\circ$ in the N- and F-regions, suggesting that only modest local geometric adjustments are required to accommodate the orientational mismatch between the two grains.

In contrast, for the high-energy $[0\bar{1}\bar{1}]$ TNS, the geometric distortions are no longer confined to the I-region but extend into the N- and F-regions, with larger regional deviations in the strut length, strut angle, dihedral angle, and node coplanarity than those of the low-energy GBs. More importantly, although the $[100]$ ATNS contains no new nodes, its coplanarity deviations in the N- and F-regions reach $11.85^\circ$ and $13.47^\circ$, respectively, which are the largest among all GBs considered. Meanwhile, its strut length, strut angle, and dihedral angle also exhibit relatively large geometric distortions, indicating that the distortions extends  extend over a much broader region rather than remaining localized near a reconstructed region. Such widely distributed distortions impose bending and orientational constraints on polymer chains, thereby reducing their conformational freedom and increasing the entropy penalty.

These results demonstrate that GB stability is governed not only by the formation of new nodes but also by the spatial distribution of geometric distortions within the network. Variations in dihedral angle, strut length, and strut angle provide relatively flexible degrees of freedom for accommodating geometric mismatch, whereas node coplanarity should be preserved whenever possible. When the mismatch is primarily accommodated through the former, a relatively low GB energy can still be maintained even in the presence of locally reconstructed nodes. By contrast, even in the absence of new-node formation, widespread disruption of node coplanarity throughout the network results in a substantial entropy penalty and a high GB energy. Therefore, preserving node coplanarity and avoiding new-node formation alone are the key geometric requirement for achieving a low-energy GB configuration in DG networks.

\section{Conclusion}

In this work, we have developed a SCFT framework designed for investigating GBs in ordered phases of BCPs, as well as an accurate and efficient numerical method for computing GB energies. 
Using this framework, we construct two energy landscapes for tilt and twist GBs in the DG phase by sampling a total of 33 same orientation points in the $(\theta,\Psi)$ space.
The two energy landscapes reveal multiple local minima, among which an asymmetric tilt network-switching GB ($[\bar{1}\bar{1}1]$ ATNS) is identified as the lowest-energy GB, with a lower energy than the experimentally reported $[1\bar{2}\bar{1}]$ TNS. Guided by these energy landscapes, we select four representative network-switching GBs for detailed investigations, including the low-energy $[\bar{1}\bar{1}1]$ ATNS, $[1\bar{2}\bar{1}]$ TNS, together with higher-energy twist ($[0\bar{1}\bar{1}]$ TNS) and tilt ($[100]$ ATNS) GBs for comparative analyses. 
GB energy decomposition demonstrates that the stability of GBs in BCP network phases is predominantly governed by the entropy contribution.

Detailed analyses of chain trajectories, IMDS geometry, and network topology elucidate the microscopic origin of this stability. 
For twist-type GBs, the formation of new nodes and the disruption of node coplanarity impose strong conformational and orientational constraints on polymer chains, leading to severe packing frustration and large entropy penalties. Notably, although the $[100]$ ATNS contains no new nodes, its extensive disruption of node coplanarity still gives rise to a large entropy penalty. In contrast, the $[\bar{1}\bar{1}1]$ ATNS connects two grains via existing struts, avoiding new-node generation and largely preserving node coplanarity. This allows geometric mismatch to be accommodated through relaxed conformation adjustment, thereby minimizing entropy penalty and yielding superior thermodynamic stability.
These findings reveal an entropy-driven physical mechanism for GB stability in soft-matter network phases. Low-energy GBs are selected by minimizing entropy penalties, notably the avoidance of new-node formation and the preservation of nodal coplanarity, rather than by strict lattice matching.
More broadly, this work provides a theoretical framework and guiding principles for understanding and engineering defects in complex soft-matter networks, offering insights into a wide class of self-assembled materials.

\section{Associated content}
\textbf{Supporting Information}

Energy landscape construction method, Two illustrative GB constructions.

Figures\,S1-S3 (PDF)

\section{Author information}
	\textbf{Corresponding authors}
	
	\textbf{Dan Wei} - \textit{ Hunan Key Laboratory for Computation and Simulation in Science and Engineering, Key Laboratory of Intelligent Computing and Information Processing of Ministry of Education, School of Mathematics and Computational Science, Xiangtan University, Xiangtan, Hunan, China, 411105}
	
	E-mail: danwei@xtu.edu.cn 
	
	\textbf{An-Chang Shi} - \textit{Department of Physics and Astronomy, McMaster University, Hamilton, Ontario, L8S4M1, Canada}
	
	E-mail: shi@mcmaster.ca
	
	\textbf{Kai Jiang} - \textit{ Wuhan Institute for Math \& AI, Wuhan University, Wuhan, Hubei, China, 430072}
	
	E-mail: kaijiang@whu.edu.cn 

	\hspace*{\fill}
		
	\noindent \textbf{Authors}
	
	\textbf{Jing Chen} - \textit{Hunan Key Laboratory for Computation and Simulation in Science and Engineering, Key Laboratory of Intelligent Computing and Information Processing of Ministry of Education, School of Mathematics and Computational Science, Xiangtan University, Xiangtan, Hunan, China, 411105.}
	
	\textbf{Aiping Zhu} - \textit{Hunan Key Laboratory for Computation and Simulation in Science and Engineering, Key Laboratory of Intelligent Computing and Information Processing of Ministry of Education, School of Mathematics and Computational Science, Xiangtan University, Xiangtan, Hunan, China, 411105}
	
	\hspace*{\fill}
		
	\noindent \textbf{Notes} \\
	The authors declare no competing financial interest.

\begin{acknowledgement}
This work is supported by the National Key R\&D Program of China (2023YFA1008802), the Science and Technology Innovation Program of Hunan Province (2024RC1052), the Innovative Research Group Project of Natural Science Foundation of Hunan Province of China (2024JJ1008), the Natural Sciences and Engineering Research Council of Canada, the Postgraduate Scientific Research Innovation Project of Hunan Province (CX20240589), the Postgraduate Scientific Research Innovation Project of Xiangtan University (XDCX2025Y237), and the High Performance Computing Platform of Xiangtan University.
\end{acknowledgement}

% \begin{suppinfo}

% % The following files are available free of charge.
% % \begin{itemize}
% %   \item Filename: brief description
% %   \item Filename: brief description
% % \end{itemize}

% \end{suppinfo}

\bibliography{achemso-demo}

\end{document}

%\begin{tocentry}
%
%Some journals require a graphical entry for the Table of Contents.
%This should be laid out ``print ready'' so that the sizing of the
%Inside the \texttt{tocentry} environment, the font used is Helvetica
%8\,pt, as required by \emph{Journal of the American Chemical
%The surrounding frame is 9\,cm by 3.5\,cm, which is the maximum
%permitted for  \emph{Journal of the American Chemical Society}
%graphical table of content entries. The box will not resize if the
%content is too big: instead it will overflow the edge of the box.
%
%This box and the associated title will always be printed on a
%separate page at the end of the document.
%
%\end{tocentry}



%\renewcommand*{\thetable}{S\arabic{table}}

\begin{figure}[htbp]
    \centering
    \includegraphics[width=0.92\linewidth]{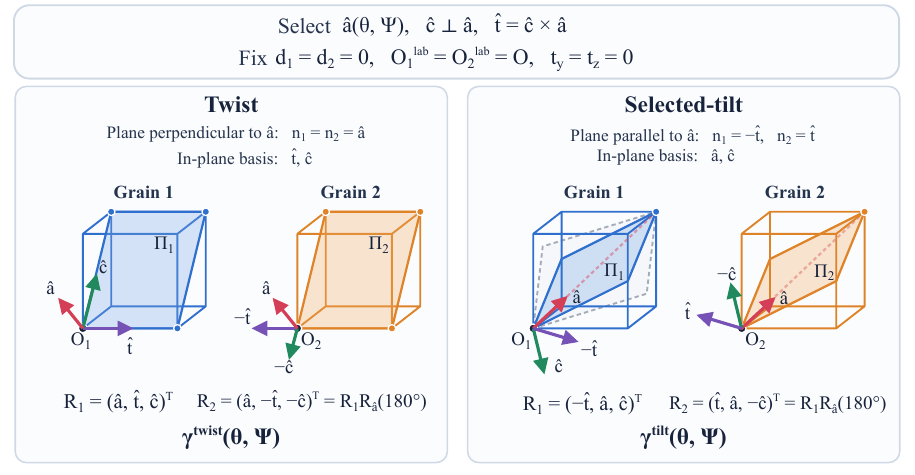}
    \caption{Construction of the twist and selected-tilt GB configurations. For each rotation axis $\hat{\mathbf a}(\theta,\Psi)$, the relative rotation is fixed at $180^{\circ}$, the cutoff planes are selected to pass through the respective unit-cell origins, giving $d_1=d_2=0$, and the relative in-plane translation is fixed at $t_y=t_z=0$. An auxiliary crystallographic direction $\hat{\mathbf c}$ perpendicular to $\hat{\mathbf a}$ is determined according to the prescribed rule, and $\hat{\mathbf t}=\hat{\mathbf c}\times\hat{\mathbf a}$. For the twist configuration, the selected cutoff plane is perpendicular to $\hat{\mathbf a}$, and the rotation matrices of the two grains are $R_1^{\mathrm{twist}} =(\hat{\mathbf a},\hat{\mathbf t},\hat{\mathbf c})^{\mathrm T}$ and $R_2^{\mathrm{twist}} =(\hat{\mathbf a},-\hat{\mathbf t},-\hat{\mathbf c})^{\mathrm T}$. For the selected-tilt configuration, one representative cutoff plane containing both $\hat{\mathbf a}$ and $\hat{\mathbf c}$ is selected from the multiple planes containing the same rotation axis, with $R_1^{\mathrm{tilt}} =(-\hat{\mathbf t},\hat{\mathbf a},\hat{\mathbf c})^{\mathrm T}$ and $R_2^{\mathrm{tilt}} =(\hat{\mathbf t},\hat{\mathbf a},-\hat{\mathbf c})^{\mathrm T}$. In each rotation matrix, the first row specifies the cutoff-plane normal, while the second and third rows specify two orthogonal in-plane crystallographic directions, which are aligned with the laboratory $x$, $y$, and $z$ directions, respectively. The rotation matrices satisfy $R_1^{\mathrm T}R_2 =R_{\hat{\mathbf a}(\theta,\Psi)}(180^\circ)$. The two constructions at a given $(\theta,\Psi)$ correspond to distinct GB configurations, whose energies $\gamma^{\mathrm{twist}}(\theta,\Psi)$ and $\gamma^{\mathrm{tilt}}(\theta,\Psi)$ are evaluated independently.}
    \label{fig:gb_two_branch_8d_to_2d}
\end{figure}

\setlength{\parindent}{2em}
\noindent
This PDF file includes:\\
\ref{SI:landscape} Energy landscape construction method\\
\ref{SI:example} Two illustrative GB constructions\\
% \ref{SI:configuration-energy-landscapes_point} Sampling of the rotation-axis directions\\
\indent Figures \ref{fig:gb_eight_dof_schematic}-\ref{fig:E_landscape_point}\\

\newpage

\section{Energy landscape construction method}
\label{SI:landscape}

\subsection{General GB configuration space}
In general, a GB configuration can be constructed by selecting one crystallographic cutoff plane from the reference unit cell of each grain. For Grain $i$, the orientation of the selected cutoff plane is determined by its unit normal $\mathbf n_i$, which can be parameterized by two angles $(\theta_i,\Psi_i)$. For a given plane normal, there is a family of parallel planes with the same orientation that differ in their distances from the unit-cell origin $\mathbf O_i$. An additional parameter $d_i$ is therefore introduced to select a specific cutoff plane from this family. Thus, the cutoff plane in each grain is uniquely specified by three parameters $(\theta_i,\Psi_i,d_i)$.

Specifically, the selected cutoff plane in Grain $i$ can be written as
\begin{equation}
    \Pi_i
    =
    \left\{
    \mathbf r_i
    \;\middle|\;
    \mathbf n_i\cdot
    (\mathbf r_i-\mathbf O_i)=d_i
    \right\},
    \qquad i\in\{1,2\},
    \label{eq:general-cutoff-planes1}
\end{equation}
where $\mathbf O_i$ denotes the crystallographic origin of the reference unit cell, $\mathbf n_i=\mathbf n_i(\theta_i,\Psi_i)$ is the unit normal of the selected cutoff plane, and $d_i$ denotes the signed distance from the unit-cell origin to the cutoff plane along the unit normal $\mathbf n_i$.
\begin{figure}[htbp]
    \centering
    \includegraphics[width=0.78\linewidth]{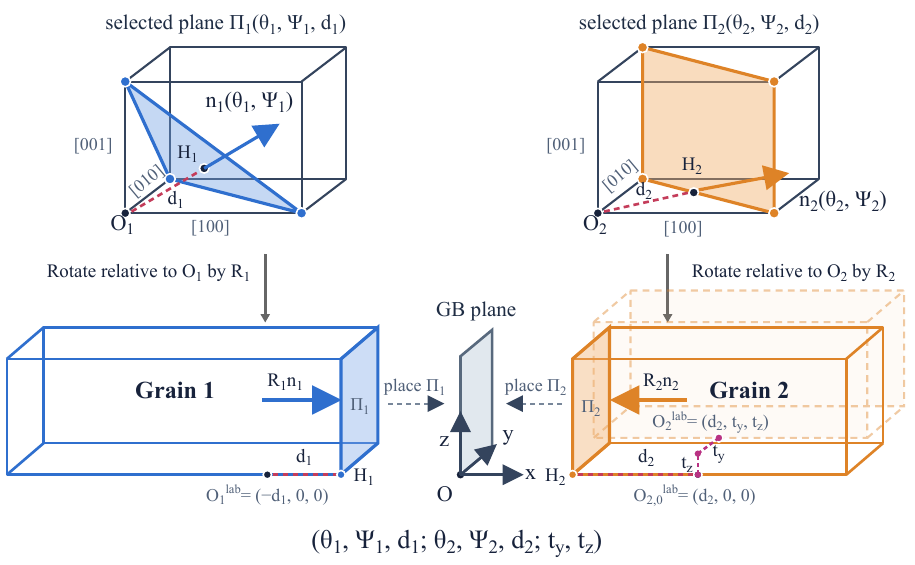}
    \caption{
    Schematic of the general GB construction with eight degrees of freedom. A cutoff plane $\Pi_i$ is selected in each reference unit cell by specifying its normal direction $\mathbf n_i(\theta_i,\Psi_i)$ and signed distance $d_i$ from the unit-cell origin $O_i$. The schematic illustrates the case $d_1,d_2>0$. Each grain is then rotated about its respective origin $O_i$ by $R_i$, such that the rotated cutoff plane is parallel to the $y$--$z$ plane. The two rotated cutoff planes are placed at $x=0$ and brought into coincidence to form the GB plane. Taking Grain~1 as the reference, Grain~2 can be translated within the GB plane by $(t_y,t_z)$ to specify the relative in-plane position of the two grains. The orange dashed rectangle represents the position of Grain 2 after the in-plane translation. Here, $H_i$ denotes the intersection of the cutoff plane $\Pi_i$ with the line passing through $O_i$ along its normal direction $\mathbf n_i$.
    }
    \label{fig:gb_eight_dof_schematic}
\end{figure}

After one cutoff plane has been selected from each grain, the two planes are brought into coincidence to form a common GB plane. The relative in-plane displacement between the two grains can still vary through two independent in-plane translations, denoted by $(t_y,t_z)$. Therefore, the geometric construction of a GB can be parameterized by
\begin{equation}
    (\theta_1,\Psi_1,d_1;\,
     \theta_2,\Psi_2,d_2;\,
     t_y,t_z),
    \label{eq:eight-dimensional-parameterization}
\end{equation}
which defines an eight-dimensional GB configuration space considered here, as illustrated schematically in Figure\,\ref{fig:gb_eight_dof_schematic}.

To place the selected cutoff planes in the SCFT simulation box, the orientations of the two grains must also be specified. Although $\mathbf n_i$ and $d_i$ determine the geometric plane $\Pi_i$, the orientation of the grain about $\mathbf n_i$ is not fixed by the plane geometry. We therefore introduce two mutually orthogonal crystallographic unit directions $\mathbf p_i$ and $\mathbf q_i$ within the selected cutoff plane, satisfying
\begin{equation}
    \mathbf p_i\cdot\mathbf n_i=0,
    \qquad
    \mathbf q_i\cdot\mathbf n_i=0,
    \qquad
    \mathbf p_i\cdot\mathbf q_i=0.
    \label{eq:inplane_basis}
\end{equation}
Together with the cutoff-plane normal, these directions determine the orientation of the grain. Importantly, $\mathbf p_i$ and $\mathbf q_i$ do not define a different geometric cutoff plane; rather, they specify two crystallographic directions within the selected cutoff plane.

For convenience in the SCFT calculations, we choose the GB plane to be located at $x=0$. Rigid-body rotations $R_1$ and $R_2$ about the respective crystallographic origins $O_1$ and $O_2$ are applied to the two grains. The rotation matrix $R_i$ maps the cutoff-plane normal and the two associated in-plane directions onto the laboratory coordinate axes according to
\begin{equation}
    R_i\mathbf n_i=\mathbf e_x,
    \qquad
    R_i\mathbf p_i=\mathbf e_y,
    \qquad
    R_i\mathbf q_i=\mathbf e_z,
    \qquad i\in\{1,2\},
    \label{eq:general_rotation_mapping}
\end{equation}
where $(\mathbf e_x,\mathbf e_y,\mathbf e_z)$ denotes the orthonormal basis of the laboratory frame. Equivalently, the corresponding rotation matrix can be written as
\begin{equation}
    R_i
    =
    \left(
        \mathbf n_i,\,
        \mathbf p_i,\,
        \mathbf q_i
    \right)^T.
    \label{eq:general_rotation_matrix}
\end{equation}
Thus, $R_i$ contains both the normal direction of the selected cutoff plane and the two crystallographic directions within the plane. The mapping of $\mathbf n_i$ onto $\mathbf e_x$ makes the cutoff plane parallel to the $y$--$z$ plane, while the mappings of $\mathbf p_i$ and $\mathbf q_i$ onto $\mathbf e_y$ and $\mathbf e_z$, respectively, fix two crystallographic directions within the selected cutoff plane.

The two rotated cutoff planes are then placed at $x=0$ and brought into coincidence to form the GB plane. Since the two in-plane basis directions of each cutoff plane have already been aligned with the common laboratory $y$ and $z$ directions through $R_1$ and $R_2$, no additional relative in-plane rotation is introduced when the two cutoff planes are brought into coincidence. Taking Grain~1 as the reference, the remaining relative placement of Grain~2 is specified by an in-plane translation. Here, $t_y$ and $t_z$ denote the relative translations along the laboratory $y$- and $z$-directions, respectively. Thus, $(t_y,t_z)$ characterizes the relative in-plane position of the two grains at the GB.

% For convenience in the SCFT calculations, we choose the GB plane to be located at $x=0$. Rigid-body rotations $R_1$ and $R_2$ about their respective crystallographic origins $O_1$ and $O_2$ are applied to Grains 1 and 2, respectively, such that the transformed normals $R_1\mathbf n_1$ and $R_2\mathbf n_2$ are parallel to the $x$-direction. \Note{In addition to aligning the cutoff-plane normals, the choice of $R_1$ and $R_2$ specifies the relative orientation of the two grains.} The corresponding cutoff planes are therefore parallel to the $y$--$z$ plane. These rigid-body rotations only change the placement of the grains and their cutoff planes in the laboratory frame and do not alter the crystallographic geometry of the selected cutoff planes.

% The two rotated cutoff planes are then placed at $x=0$ and brought into coincidence to form the GB plane. Taking Grain~1 as the reference and fixing its cutoff plane, the in-plane translation between the two grains is specified by translating Grain~2 relative to Grain~1 within the GB plane. Here, $t_y$ and $t_z$ denote the relative translations along the $y$- and $z$-directions, respectively. Thus, $(t_y,t_z)$ characterizes the relative in-plane position of the two grains at the GB.

\subsection{Reduction to two-dimensional GB configuration spaces}

In the present work, we consider two prescribed two-dimensional subsets of the general eight-dimensional GB configuration space. To obtain a computationally tractable sampling space, we use the relative orientation between the two grains to parameterize the sampled GB configurations, since varying the relative orientation intuitively changes the relative geometric arrangement of the two DG networks across the GB.
%\Note{because varying the relative orientation provides a systematic way to compare different geometric relationships between the two DG networks while the other construction variables are controlled.}
% we focus on the relative orientation between the two grains, \Noter{because variations in relative orientation provide a direct way to generate and characterize distinct geometric relationships between the two DG network structures across the GB}. 
We fix the misorientation angle at $180^\circ$ and use the direction of the relative rotation axis, parameterized by $(\theta,\Psi)$, as the two independent variables.
The remaining degrees of freedom are fixed through the prescribed geometric construction described below.

First, both grains are constructed from the same DG unit cell. For the positions of the cutoff planes relative to the unit-cell origin, we consider only the planes passing through their respective crystallographic origins $O_i$, and therefore set
\begin{equation}
    d_1=d_2=0.
\end{equation}
Meanwhile, the relative in-plane translation between the two grains is fixed to zero,
\begin{equation}
    t_y=t_z=0.
\end{equation}
Thus, after the two cutoff planes are placed at $x=0$, their crystallographic origins coincide at the same point $O$ on the GB plane ($O_1^{\mathrm{lab}}=O_2^{\mathrm{lab}}=O$). We further fix the misorientation angle between the two grains at $180^\circ$ and use the direction of the relative-rotation axis as the remaining independent variable. The rotation axis is represented by a unit vector $\hat{\mathbf a}$ passing through the unit-cell origin, with its direction parameterized by the two angles $(\theta,\Psi)$.

For a given rotation axis $\hat{\mathbf a}$, we introduce an auxiliary unit vector $\hat{\mathbf c}$ perpendicular to $\hat{\mathbf a}$. The direction $\hat{\mathbf c}$ is uniquely determined from $\hat{\mathbf a}$ according to the prescribed rule described below. We further define
\begin{equation}
    \hat{\mathbf t}
    =
    \hat{\mathbf c}\times\hat{\mathbf a},
\end{equation}
such that $\hat{\mathbf a}$, $\hat{\mathbf c}$, and $\hat{\mathbf t}$ form an orthogonal basis. Based on this auxiliary orthogonal basis, we construct two families of GB configurations, namely the twist and selected-tilt families.

For the \textbf{twist family}, the selected cutoff planes are perpendicular to the rotation axis $\hat{\mathbf a}$. In Grain~1, the cutoff plane is defined by the two in-plane directions $\hat{\mathbf t}$ and $\hat{\mathbf c}$, with its normal given by $\hat{\mathbf a}$. The orientation of Grain~2 relative to Grain~1 is specified by a $180^\circ$ rotation about $\hat{\mathbf a}$.
Accordingly, the two in-plane directions become $-\hat{\mathbf t}$ and $-\hat{\mathbf c}$, whereas the plane normal $\hat{\mathbf a}$ remains unchanged. To orient the two cutoff planes parallel to the $x=0$ plane, the following rigid-body rotations are applied to the two same DG unit cells,
\begin{equation}
    R_1^{\mathrm{twist}}
    =
    \left(
    \hat{\mathbf a},
    \hat{\mathbf t},
    \hat{\mathbf c}
    \right)^{T},
    \qquad
    R_2^{\mathrm{twist}}
    =
    \left(
    \hat{\mathbf a},
    -\hat{\mathbf t},
    -\hat{\mathbf c}
    \right)^{T}.
    \label{eq:rotation_twist}
\end{equation}
The two rotation matrices satisfy
\begin{equation}
    R_2^{\mathrm{twist}}
    =
    R_1^{\mathrm{twist}}
    R_{\hat{\mathbf a}}(180^\circ),
\end{equation}
and preserve the prescribed $180^\circ$ relative orientation between the two grains. Consequently, once
$\hat{\mathbf a}(\theta,\Psi)$ is specified, the orientations of the two cutoff planes in the twist family are fully determined.

For the \textbf{selected-tilt family}, the rotation axis $\hat{\mathbf a}$ lies within the selected cutoff plane. For a given rotation axis, there are multiple planes containing $\hat{\mathbf a}$. To select a specific plane, we use the auxiliary direction $\hat{\mathbf c}$, and select the plane containing both $\hat{\mathbf a}$ and $\hat{\mathbf c}$. In Grain~1, this plane is defined by $\hat{\mathbf a}$ and $\hat{\mathbf c}$, with its normal given by $-\hat{\mathbf t}$. The orientation of Grain~2 relative to Grain~1 is specified by a $180^\circ$ rotation about $\hat{\mathbf a}$. Under this rotation, $\hat{\mathbf a}$ remains unchanged, whereas $\hat{\mathbf c}$ becomes $-\hat{\mathbf c}$, so that the corresponding plane normal in Grain~2 is $\hat{\mathbf t}$. To orient the two grains so that their cutoff planes coincide with the $x=0$ plane, we apply
\begin{equation}
    R_1^{\mathrm{tilt}}
    =
    \left(
    -\hat{\mathbf t},
    \hat{\mathbf a},
    \hat{\mathbf c}
    \right)^{T},
    \qquad
    R_2^{\mathrm{tilt}}
    =
    \left(
    \hat{\mathbf t},
    \hat{\mathbf a},
    -\hat{\mathbf c}
    \right)^{T}.
    \label{eq:rotation_tilt}
\end{equation}
These rotation matrices likewise satisfy
\begin{equation}
    R_2^{\mathrm{tilt}}
    =
    R_1^{\mathrm{tilt}}
    R_{\hat{\mathbf a}}(180^\circ).
\end{equation}
Therefore, the normals of the two selected cutoff planes in the selected-tilt family are $-\hat{\mathbf t}$ and $\hat{\mathbf t}$, respectively. Since
\begin{equation}
    \hat{\mathbf t}
    =
    \hat{\mathbf c}\times\hat{\mathbf a},
\end{equation}
and $\hat{\mathbf c}$ is uniquely determined from $\hat{\mathbf a}$, the orientations of the two cutoff planes in the selected-tilt family are ultimately determined by $\hat{\mathbf a}(\theta,\Psi)$ as well.

With the above construction, all the degrees of freedom in the eight-dimensional GB parameterization
\begin{equation}
    (\theta_1,\Psi_1,d_1;\,
     \theta_2,\Psi_2,d_2;\,
     t_y,t_z)
\end{equation}
are either determined by $(\theta,\Psi)$ or fixed by the prescribed constraints. Specifically, for the twist family, the normals of the two cutoff planes are directly determined by the rotation axis $\hat{\mathbf a}(\theta,\Psi)$. For the selected-tilt family, the plane normals are determined by $\pm\hat{\mathbf t}$, where $\hat{\mathbf t}$ is uniquely determined from $\hat{\mathbf a}$ through the prescribed auxiliary direction $\hat{\mathbf c}$. In addition, $d_1=d_2=0$ and $t_y=t_z=0$ are fixed throughout the construction. Therefore, for each $(\theta,\Psi)$, the above construction uniquely determines one twist configuration and one selected-tilt configuration, resulting in two two-dimensional families parameterized by $(\theta,\Psi)$ within the eight-dimensional GB configuration space. We emphasize that these two families represent prescribed subsets of the full eight-dimensional space. In particular, for each rotation axis, the prescribed construction selects only one tilt plane from the multiple planes containing the same rotation axis and therefore does not include all possible tilt configurations.

\subsection{SCFT calculation of the GB energy landscapes}

\begin{figure}[htbp]
	\centering
	\includegraphics[width=1\linewidth]{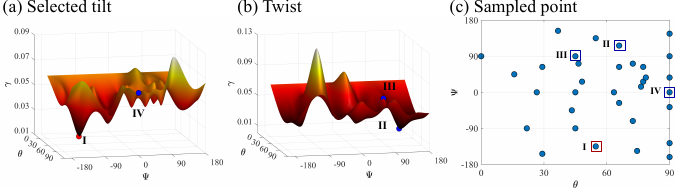}
	\caption{
    GB energy landscapes and sampled rotation-axis directions. (a) Energy landscape of the selected-tilt GB family.
    (b) Energy landscape of the twist GB family. (c) Distribution of the sampled rotation-axis directions in the $(\theta,\Psi)$ parameter space, with $\theta\in[0^\circ,90^\circ]$ and $\Psi\in[-180^\circ,180^\circ]$.
    The energy landscapes in (a) and (b) are obtained by interpolation of the discrete GB energies calculated by SCFT at the sampled points. The labeled points I--IV indicate the representative GB configurations
    discussed in the main text.}
	\label{fig:E_landscape_point}
\end{figure}

Based on the two-dimensional GB configuration spaces described above, we calculate the energy landscapes of the twist and selected-tilt GB families by sampling the direction of the relative-rotation axis $\hat{\mathbf a}(\theta,\Psi)$. As shown in Figure\,\ref{fig:E_landscape_point}c, a total of 33 rational rotation-axis directions are sampled in the $(\theta,\Psi)$ parameter space. For each sampling point, one twist GB and one selected-tilt GB are constructed according to the
geometric rules described in the preceding section, and the two configurations are independently calculated by SCFT. Therefore, the 33 sampled rotation-axis directions correspond to a total of 66 independent GB calculations.

For a given rotation axis $\hat{\mathbf a}(\theta,\Psi)$, the rotation matrices $R_1$ and $R_2$ for the two grains in the twist or selected-tilt configuration are first determined according to the constructions described above. For Grain $m$, the A-block segment density after applying the rotation $R_m$ to the bulk DG structure is given by
\begin{equation}
    \phi_{\mathrm{A},\,\mathrm{grain}}^{R_m}(\br)
    =
    \phi_{\mathrm{A},\,\mathrm{bulk}}
    \left(R_m^T\br\right)
    =
    \sum_{\bk\in\mathbb{Z}^{3}}
    \hat{\phi}_{\mathrm{A},\,\mathrm{bulk}}(\bk)
    e^{i\left(R_m\mathcal{B}\bk\right)^T\br},
    \qquad m\in\{1,2\},
    \label{eq:rotated_bulk_density}
\end{equation}
where $\mathcal{B}$ denotes the primitive reciprocal lattice and $\hat{\phi}_{\mathrm{A},\,\mathrm{bulk}}(\bk)$ denotes the Fourier coefficient of the A-block segment density of the bulk DG phase. This procedure gives the A-block segment densities of the rotated Grain 1 and Grain 2.

The A-block segment densities of the two rotated grains are then smoothly connected in the vicinity of $x=0$ to construct the initial A-block segment density of the GB system,
\begin{equation}
    \phi_{\mathrm{A},\,\mathrm{GB}}(\br)
    =
    [1-c(x)]
    \phi_{\mathrm{A},\,\mathrm{grain}}^{R_1}(\br)
    +
    c(x)
    \phi_{\mathrm{A},\,\mathrm{grain}}^{R_2}(\br),
    \label{eq:GB_initial_density}
\end{equation}
where the smooth function is defined as
\begin{equation}
    c(x)
    =
    \frac{1+\tanh(\sigma x)}{2},
    \label{eq:smooth_function}
\end{equation}
with $\sigma$ controlling the width of the transition region between the two grains near the GB. The initial B-block segment density is determined from the incompressibility condition. In this way, the initial SCFT configurations of the twist and selected-tilt GBs are constructed for each sampled $(\theta,\Psi)$.

Starting from these initial segment densities, each GB configuration is independently relaxed by SCFT until convergence. The GB energy of the converged configuration is calculated as
\begin{equation}
    \gamma^s\bigl((\theta,\Psi)_j\bigr)
    =
    \frac{L}{n k_{\mathrm B}T}
    \left[
        F_{\mathrm{GB}}^s\bigl((\theta,\Psi)_j\bigr)
        -
        F_{\mathrm{grain}}\bigl((\theta,\Psi)_j\bigr)
    \right],
    \quad
    s\in\{\mathrm{twist},\mathrm{tilt}\},
    \quad j=1,\ldots,33.
\end{equation}
where $F_{\mathrm{GB}}^s/nk_{\mathrm B}T$ and $F_{\mathrm{grain}}/nk_{\mathrm B}T$ are the free-energy densities of the converged GB system and the corresponding bulk grain, respectively, and $L$ is the simulation-box length in the $x$ direction.

Repeating the above procedure for all 33 sampled points $(\theta,\Psi)_j$, $j=1,\ldots,33$, gives the two sets of discrete
GB energies
\begin{equation}
    \left\{
    \gamma^{s}\bigl((\theta,\Psi)_j\bigr)
    \right\}_{j=1}^{33},
    \qquad
    s\in\{\mathrm{twist},\mathrm{tilt}\}.
    \label{eq:discrete_GB_energies}
\end{equation}
The GB energy at each sampling point is obtained directly from an independent converged SCFT calculation. The two sets of discrete GB energies are then interpolated separately over the $(\theta,\Psi)$ parameter space to obtain the corresponding two-dimensional GB energy landscapes, $\gamma^{\mathrm{twist}}(\theta,\Psi)$ and $\gamma^{\mathrm{tilt}}(\theta,\Psi)$. Figure\,\ref{fig:E_landscape_point}a-b shows the resulting selected-tilt and twist GB energy landscapes. Since inversion-related rotation axes generate equivalent GB configurations, the two energy landscapes are shown only for the region $(\theta,\Psi)\in[0^\circ,90^\circ]\times[-180^\circ,180^\circ)$.

\subsection{Parameterization of the relative orientation}
\label{SI:relative orientation}

Let $R_i$ denote the orientation matrix that maps the crystallographic coordinates of Grain~$i$ to a common reference frame. The relative orientation of Grain 2 with respect to Grain 1 is defined as
\begin{equation}
    \Delta R
    =
    R_1^{\mathrm T}R_2.
    \label{eq:relative-orientation-definition}
\end{equation}
Taking Grain~1 as the reference, the relative orientation can be represented by the equivalent orientation pair $(I,\Delta R)$. This representation specifies only the relative crystallographic orientation of the two grains. It is unchanged by any rigid-body rotation or translation subsequently used to place the two-grain configuration in the laboratory frame.

In the axis--angle representation, the relative rotation is written as
\begin{equation}
    \Delta R
    =
    R_{\hat{\mathbf a}}(\omega)
    =
    \cos\omega\,I
    +
    (1-\cos\omega)
    \hat{\mathbf a}\hat{\mathbf a}^{\mathrm T}
    +
    \sin\omega\,[\hat{\mathbf a}]_{\times},
\end{equation}
where $I$ is the identity matrix and $[\hat{\mathbf a}]_{\times}$ is the skew-symmetric matrix associated with the rotation axis $\hat{\mathbf a}$. The direction of the unit rotation axis is parameterized by the spherical coordinates
$(\theta,\Psi)$ as
\begin{equation}
    \hat{\mathbf a}(\theta,\Psi)
    =
    \begin{pmatrix}
        \sin\theta\cos\Psi\\
        \sin\theta\sin\Psi\\
        \cos\theta
    \end{pmatrix},
    \label{eq:a(theta-phi)}
\end{equation}
where $\theta$ and $\Psi$ denote the polar and azimuthal angles, respectively.

In the present work, the sampled rotation axes are chosen along rational crystallographic directions $[hkl]$, where $h$, $k$, and $l$ are reduced to coprime integers. The corresponding spherical angles are
\begin{equation}
    \theta
    =
    \arccos
    \frac{l}{\sqrt{h^2+k^2+l^2}},
    \qquad
    \Psi
    =
    \operatorname{atan2}(k,h).
    \label{eq:hkl-to-angles}
\end{equation}

We fix the misorientation angle at $\omega=180^\circ$. Under this condition, the relative orientation is fully characterized by the direction of the rotation axis $\hat{\mathbf a}$, or equivalently by $(\theta,\Psi)$. The relative orientation matrix therefore reduces to
\begin{equation}
    \Delta R(\theta,\Psi)
    =
    R_{\hat{\mathbf a}(\theta,\Psi)}(180^\circ)
    =
    2\hat{\mathbf a}(\theta,\Psi)
    \hat{\mathbf a}^{\mathrm T}(\theta,\Psi)
    -
    I.
    \label{eq:deltaR}
\end{equation}

Thus, each sampling point $(\theta,\Psi)$ determines one relative orientation $\Delta R(\theta,\Psi)$. At $\omega=180^\circ$, the reverse axes $\hat{\mathbf a}$ and $-\hat{\mathbf a}$ generate the same relative orientation matrix, so only one of the two equivalent directions is retained in the sampling domain. Importantly, $(\theta,\Psi)$ specifies only the relative orientation of the two grains and does not by itself determine a complete GB configuration.

\subsection{Construction of auxiliary direction $\hat{\mathbf c}$}
\label{SI:boundary plane}
For the GB construction described above, an auxiliary crystallographic direction $\hat{\mathbf c}$ perpendicular to the rotation axis $\hat{\mathbf a}$ is required to specify a cutoff plane. Since multiple directions can be perpendicular to the same rotation axis, we adopt the following prescribed rule to determine $\hat{\mathbf c}$ consistently for each sampled axis.

For a sampled rational rotation axis $[hkl]$, we first identify all primitive integer directions perpendicular to the axis. The corresponding candidate set is defined as
\begin{equation}
\mathcal C(h,k,l)
=
\left\{
(u,v,w)\in\mathbb Z^3\setminus\{\mathbf 0\}
\;\middle|\;
hu+kv+lw=0,\;
\gcd(|u|,|v|,|w|)=1
\right\}.
\end{equation}
Here, $hu+kv+lw=0$ ensures that each candidate direction is perpendicular to the rotation axis, while $\gcd(|u|,|v|,|w|)=1$ removes integer multiples of the same crystallographic direction. 

We then select a direction from this candidate set. As the first criterion, we choose the shortest primitive crystallographic direction, i.e., the direction with the smallest Euclidean length. We first define
\begin{equation}
\mathcal D_{\min}(h,k,l)
=
\operatorname*{min}_{(u,v,w)\in\mathcal C(h,k,l)}
\left(
u^2+v^2+w^2
\right),
\end{equation}
and collect all directions having this minimum length as
\begin{equation}
\mathcal C_{\min}(h,k,l)
=
\left\{
(u,v,w)\in\mathcal C(h,k,l)
\;\middle|\;
u^2+v^2+w^2=\mathcal D_{\min}(h,k,l)
\right\}.
\end{equation}
Therefore, this first step reduces the possible auxiliary directions to the shortest primitive crystallographic directions perpendicular to $\hat{\mathbf a}$.

In some cases, more than one crystallographic direction has the same minimum length.
% To remove this remaining ambiguity and make the construction reproducible, we apply a fixed tie-breaking rule. 
To select one of them uniquely and make the construction reproducible, we apply the following fixed selection rule. Among the shortest candidates, we successively minimize $|w|$, $|v|$, and $|u|$. Specifically,
\begin{equation}
\mathcal C_w(h,k,l)
=
\operatorname*{arg\,min}_{(u,v,w)\in\mathcal C_{\min}(h,k,l)}
|w|,
\end{equation}
followed by
\begin{equation}
\mathcal C_v(h,k,l)
=
\operatorname*{arg\,min}_{(u,v,w)\in\mathcal C_w(h,k,l)}
|v|,
\end{equation}
and finally
\begin{equation}
\mathcal C_*(h,k,l)
=
\operatorname*{arg\,min}_{(u,v,w)\in\mathcal C_v(h,k,l)}
|u|.
\end{equation}
In other words, candidates with the smallest $|w|$ are retained first; if more than one candidate remains, those with the smallest $|v|$ are retained; and if necessary, $|u|$ is used as the final criterion.

After these successive selections, $\mathcal C_*(h,k,l)$ contains two vectors. These two vectors represent opposite crystallographic directions and lead to geometrically equivalent GB constructions. We choose either one as the representative vector,
\begin{equation}
\hat{\mathbf c}_*
=
(u_*,v_*,w_*)^{\mathrm T}
\in
\mathcal C_*(h,k,l),
\end{equation}
and define the auxiliary unit vector as
\begin{equation}
\hat{\mathbf c}
=
\frac{\hat{\mathbf c}_*}{\|\hat{\mathbf c}_*\|}
=
\frac{(u_*,v_*,w_*)^{\mathrm T}}
{\sqrt{u_*^2+v_*^2+w_*^2}}.
\end{equation}

This procedure provides a consistent auxiliary direction $\hat{\mathbf c}$ for each sampled rational rotation axis
$\hat{\mathbf a}$. Together with $\hat{\mathbf t}=\hat{\mathbf c}\times\hat{\mathbf a}$, it determines the auxiliary orthogonal basis used in the GB construction.

\section{Two illustrative GB constructions}
\label{SI:example}
With the cutoff-plane and relative in-plane translation fixed as described above, a specific GB configuration is determined once the rotation matrices of the two grains are specified. To further illustrate how the prescribed procedure determines a specific GB configuration, we present one selected-tilt construction and one twist construction. In each example, the rotation axis $\hat{\mathbf a}$ is specified first, followed by the auxiliary direction $\hat{\mathbf c}$ determined according to the rule described above. The third direction is then defined as $\hat{\mathbf t}=\hat{\mathbf c}\times\hat{\mathbf a}.$ The three mutually orthogonal directions $\hat{\mathbf a}$, $\hat{\mathbf c}$, and $\hat{\mathbf t}$ are arranged according to the prescribed selected-tilt or twist construction to form the ordered orthogonal bases of the two grains. The corresponding rotation matrices are then obtained from these ordered bases, thereby determining the orientations of the two grains in the laboratory frame.

\subsection{Selected-tilt construction for the rotation axis $[\bar{1}\bar{1}1]$}
\label{SI:example1}

We first consider the $[\bar{1}\bar{1}1]$ ATNS constructed using the selected-tilt procedure. The rotation axis passes through the coincident reference origins $O_1$ and $O_2$ along the $[\bar{1}\bar{1}1]$ direction. Its unit vector in the crystallographic coordinate
frame is
\begin{equation}
    \hat{\mathbf a}
    =
    \frac{1}{\sqrt{3}}
    (-1,-1,1)^{\mathrm T}.
    \label{eq:ATNS-axis}
\end{equation}
An integer direction $[uvw]$ perpendicular to this axis satisfies
\begin{equation}
    -u-v+w=0.
    \label{eq:ATNS-perpendicular-condition1}
\end{equation}

According to the auxiliary-direction selection rule described above, the
shortest primitive integer solutions of
Eq.\,\eqref{eq:ATNS-perpendicular-condition1} consist of three pairs related by sign reversal,
\begin{equation}
    \pm[\bar{1}10],\qquad
    \pm[101],\qquad
    \pm[011].
    \label{eq:ATNS-shortest-directions}
\end{equation}
These three pairs have the same squared length,
\begin{equation}
    u^2+v^2+w^2=2.
\end{equation}
We apply the next criterion in the prescribed selection rule and compare their values of $|w|$. The corresponding values are $0$, $1$, and $1$ for $[\bar{1}10]$, $[101]$, and $[011]$, respectively. Thus, $\pm[\bar{1}10]$ is selected. Since the two opposite directions lead to geometrically equivalent GB constructions, we choose $[\bar{1}10]$ as the representative direction, giving
\begin{equation}
    \hat{\mathbf c}
    =
    \frac{1}{\sqrt{2}}
    (-1,1,0)^{\mathrm T}.
    \label{eq:ATNS-c}
\end{equation}
The third direction is then
\begin{equation}
    \hat{\mathbf t}
    =
    \hat{\mathbf c}\times\hat{\mathbf a}
    =
    \frac{1}{\sqrt{6}}
    (1,1,2)^{\mathrm T}.
    \label{eq:ATNS-t}
\end{equation}

For the selected-tilt construction, the rotation axis $\hat{\mathbf a}$ lies within the cutoff plane. Since multiple planes can contain the same rotation axis, the auxiliary direction $\hat{\mathbf c}$ is used to select one representative plane, namely the plane containing both $\hat{\mathbf a}$ and $\hat{\mathbf c}$. For Grain 1, the normal of this plane is $-\hat{\mathbf t}$, and the corresponding ordered basis is defined as
\begin{equation}
    \mathcal E_1^{\mathrm{tilt}}
    =
    \left(
        -\hat{\mathbf t},
        \hat{\mathbf a},
        \hat{\mathbf c}
    \right).
    \label{eq:ATNS-grain1-basis}
\end{equation}
Here, the first basis vector $-\hat{\mathbf t}$ is normal to the cutoff plane, whereas $\hat{\mathbf a}$ and $\hat{\mathbf c}$ lie within the plane. Accordingly, the Grain-1 cutoff plane contains the crystallographic
directions $[\bar{1}\bar{1}1]$ and $[\bar{1}10]$ and has the normal $[\bar{1}\bar{1}\bar{2}]$.

Grain~2 is oriented relative to Grain~1 by the prescribed $180^\circ$ rotation about $\hat{\mathbf a}$. Under this rotation, $\hat{\mathbf a}$ remains unchanged, whereas the two directions perpendicular to the rotation axis, $-\hat{\mathbf t}$ and $\hat{\mathbf c}$, reverse direction. The ordered basis of Grain 2 is therefore
\begin{equation}
    \mathcal E_2^{\mathrm{tilt}}
    =
    \left(
        \hat{\mathbf t},
        \hat{\mathbf a},
        -\hat{\mathbf c}
    \right).
    \label{eq:ATNS-grain2-basis}
\end{equation}
The Grain-2 cutoff plane consequently contains the crystallographic directions $[\bar{1}\bar{1}1]$ and $[1\bar{1}0]$ and has the normal $[112]$. The $(112)$ and $(\bar{1}\bar{1}\bar{2})$ planes have opposite normal directions and therefore represent the same unoriented plane orientation. Nevertheless, the two grains have different crystallographic orientations because they remain related by the prescribed $180^\circ$ relative rotation about $\hat{\mathbf a}$.

The laboratory-frame orientation matrices of the two grains are obtained from the transposes of the corresponding ordered bases,
% \begin{equation}
%     R_1^{\mathrm{tilt}}
%     =
%     \left(
%         \mathcal E_1^{\mathrm{tilt}}
%     \right)^{\mathrm T},
%     \qquad
%     R_2^{\mathrm{tilt}}
%     =
%     \left(
%         \mathcal E_2^{\mathrm{tilt}}
%     \right)^{\mathrm T}.
%     \label{eq:ATNS-laboratory-orientations}
% \end{equation}
\begin{equation}
    R_1^{\mathrm{tilt}}
    =
    \begin{pmatrix}
        -\dfrac{1}{\sqrt{6}} & -\dfrac{1}{\sqrt{6}} & -\dfrac{2}{\sqrt{6}}
        \\[4pt]
        -\dfrac{1}{\sqrt{3}} & -\dfrac{1}{\sqrt{3}} &  \dfrac{1}{\sqrt{3}}
        \\[4pt]
        -\dfrac{1}{\sqrt{2}} &  \dfrac{1}{\sqrt{2}} & 0
    \end{pmatrix},
    \qquad
    R_2^{\mathrm{tilt}}
    =
    \begin{pmatrix}
        \dfrac{1}{\sqrt{6}} &  \dfrac{1}{\sqrt{6}} & \dfrac{2}{\sqrt{6}}
        \\[4pt]
        -\dfrac{1}{\sqrt{3}} & -\dfrac{1}{\sqrt{3}} & \dfrac{1}{\sqrt{3}}
        \\[4pt]
        \dfrac{1}{\sqrt{2}} & -\dfrac{1}{\sqrt{2}} & 0
    \end{pmatrix}.
    \label{eq:ATNS-explicit-orientation-matrices}
\end{equation}

Accordingly, the Grain-1 directions $[\bar{1}\bar{1}\bar{2}]$, $[\bar{1}\bar{1}1]$, $[\bar{1}10]$ are aligned with the laboratory $x$, $y$, and $z$ directions, respectively. For Grain 2, the corresponding directions are $[112]$, $[\bar{1}\bar{1}1]$, $[1\bar{1}0]$. With the rotation matrices of the two grains determined, the initial GB configuration with the rotation axis
$[\bar{1}\bar{1}1]$ is constructed using Eqs.\,\ref{eq:rotated_bulk_density}-\ref{eq:GB_initial_density}.

\subsection{Twist construction for the rotation axis $[1\bar{2}\bar{1}]$}
\label{SI:example2}

We next consider the twist GB with the rotation axis $[1\bar{2}\bar{1}]$. The corresponding unit vector in the
crystallographic coordinate frame is
\begin{equation}
    \hat{\mathbf a}
    =
    \frac{1}{\sqrt{6}}
    (1,-2,-1)^{\mathrm T}.
    \label{eq:TNS-axis}
\end{equation}
An integer crystallographic direction $[uvw]$ perpendicular to this axis satisfies
\begin{equation}
    u-2v-w=0.
    \label{eq:TNS-perpendicular-condition1}
\end{equation}
The shortest primitive integer solutions of Eq.\,\eqref{eq:TNS-perpendicular-condition1} are
\begin{equation}
    \pm[\bar{1}0\bar{1}],
\end{equation}
with
\begin{equation}
    u^2+v^2+w^2=2.
\end{equation}
Choosing $[\bar{1}0\bar{1}]$ as the representative direction gives
\begin{equation}
    \hat{\mathbf c}
    =
    \frac{1}{\sqrt{2}}
    (-1,0,-1)^{\mathrm T}.
    \label{eq:TNS-c}
\end{equation}
The third direction is then
\begin{equation}
    \hat{\mathbf t}
    =
    \hat{\mathbf c}\times\hat{\mathbf a}
    =
    \frac{1}{\sqrt{3}}
    (-1,-1,1)^{\mathrm T}.
    \label{eq:TNS-t}
\end{equation}

In contrast to the selected-tilt construction, the cutoff plane in the twist construction is perpendicular to the rotation axis $\hat{\mathbf a}$. Therefore, its orientation is determined directly by $\hat{\mathbf a}$, while $\hat{\mathbf t}$ and $\hat{\mathbf c}$ provide two orthogonal directions within the cutoff plane for constructing the
ordered basis.

For Grain~1, the ordered basis is defined as
\begin{equation}
    \mathcal E_1^{\mathrm{twist}}
    =
    \left(
        \hat{\mathbf a},
        \hat{\mathbf t},
        \hat{\mathbf c}
    \right).
    \label{eq:TNS-grain1-basis}
\end{equation}
Here, $\hat{\mathbf a}$ is normal to the cutoff plane, whereas $\hat{\mathbf t}$ and $\hat{\mathbf c}$ lie within the plane. Accordingly, the Grain-1 cutoff plane contains the crystallographic directions $[\bar{1}\bar{1}1]$ and $[\bar{1}0\bar{1}]$ and has the normal $[1\bar{2}\bar{1}]$.

Grain~2 is oriented relative to Grain~1 by the prescribed $180^\circ$ rotation about $\hat{\mathbf a}$. Under this rotation, $\hat{\mathbf a}$ remains unchanged, whereas $\hat{\mathbf t}$ and $\hat{\mathbf c}$ reverse direction. The ordered basis of Grain 2 is
\begin{equation}
    \mathcal E_2^{\mathrm{twist}}
    =
    \left(
        \hat{\mathbf a},
        -\hat{\mathbf t},
        -\hat{\mathbf c}
    \right).
    \label{eq:TNS-grain2-basis}
\end{equation}
Consequently, the Grain-2 cutoff plane contains the crystallographic directions $[11\bar{1}]$ and $[101]$ and has the normal $[1\bar{2}\bar{1}]$. Since the rotation axis $[1\bar{2}\bar{1}]$ is normal to the cutoff planes of both grains, the resulting configuration is of twist type.

The laboratory-frame orientation matrices of the two grains are
% \begin{equation}
%     R_1^{\mathrm{twist}}
%     =
%     \left(
%         \mathcal E_1^{\mathrm{twist}}
%     \right)^{\mathrm T},
%     \qquad
%     R_2^{\mathrm{twist}}
%     =
%     \left(
%         \mathcal E_2^{\mathrm{twist}}
%     \right)^{\mathrm T}.
%     \label{eq:TNS-laboratory-orientations}
% \end{equation}

\begin{equation}
    R_1^{\mathrm{twist}}
    =
    \begin{pmatrix}
        \dfrac{1}{\sqrt{6}}  & -\dfrac{2}{\sqrt{6}} & -\dfrac{1}{\sqrt{6}}
        \\[4pt]
        -\dfrac{1}{\sqrt{3}} & -\dfrac{1}{\sqrt{3}} &  \dfrac{1}{\sqrt{3}}
        \\[4pt]
        -\dfrac{1}{\sqrt{2}} & 0                    & -\dfrac{1}{\sqrt{2}}
    \end{pmatrix},
    \qquad
    R_2^{\mathrm{twist}}
    =
    \begin{pmatrix}
        \dfrac{1}{\sqrt{6}} & -\dfrac{2}{\sqrt{6}} & -\dfrac{1}{\sqrt{6}}
        \\[4pt]
        \dfrac{1}{\sqrt{3}} &  \dfrac{1}{\sqrt{3}} & -\dfrac{1}{\sqrt{3}}
        \\[4pt]
        \dfrac{1}{\sqrt{2}} & 0                    &  \dfrac{1}{\sqrt{2}}
    \end{pmatrix}.
    \label{eq:explicit-twist-orientation-matrices}
\end{equation}

Accordingly, the Grain-1 directions $[1\bar{2}\bar{1}]$, $[\bar{1}\bar{1}1]$, $[\bar{1}0\bar{1}]$
are aligned with the laboratory $x$, $y$, and $z$ directions, respectively. For Grain 2, the corresponding directions are $[1\bar{2}\bar{1}]$, $[11\bar{1}]$, $[101]$. With the rotation matrices of the two grains determined, the initial GB configuration with the rotation axis $[1\bar{2}\bar{1}]$ is constructed using Eqs.\,\ref{eq:rotated_bulk_density}-\ref{eq:GB_initial_density}.